\documentclass[paper]{JHEP3}
\usepackage{graphicx}
\usepackage{epsfig}
\usepackage{latexsym}
\usepackage{amssymb}
\usepackage{amsbsy}
\usepackage{enumerate}
\usepackage{bbm}
\usepackage{url}
\usepackage{wasysym}
\usepackage[nosort]{cite}
\usepackage{pslatex}
\usepackage[latin1]{inputenc}
\usepackage[T1]{fontenc}
\usepackage[hang,sf]{subfigure}
\usepackage[english]{babel}
\graphicspath{{plots/}}
\DeclareGraphicsExtensions{.eps,.ps}
\usepackage{rotating}

\newcommand{\tmtexttt}[1]{{\ttfamily{#1}}}

\def\beq{\begin{equation}}
\def\beqn{\begin{eqnarray}}
\def\eeq{\end{equation}}
\def\eeqn{\end{eqnarray}}

\newcommand\HERWIG{{\tt HERWIG}}

\newcommand\PYTHIA{{\tt PYTHIA}}

\newcommand\HELACNLO{{\tt HELAC-NLO}}

\newcommand\ptsupp{{\tt bornsuppfact}}

\def\tb{\bar t}
\def\ttb{{t\tb}}

\def\lq{\left[} 
\def\rq{\right]}

\def\({\left(} 
\def\){\right)}

\newcommand\sss{\mathchoice%
{\displaystyle}%
{\scriptstyle}%
{\scriptscriptstyle}%
{\scriptscriptstyle}%
}

\newdimen\hbigcirc
\newdimen\wbigcirc

\newdimen\figwidth

\newcommand\figleft{upper}
\newcommand\figright{lower}

\newcommand\as{\alpha_{\sss\rm S}}
\newcommand\aem{\alpha_{\sss\rm EM}}

\newcommand\pt{p_{\sss\rm T}}
\newcommand\ptrel{p_{\sss\rm T,rel}}
\newcommand\kt{k_{\sss\rm T}}

\newcommand\ptmin{{\pt^{\min}}}

\newcommand\mur{\mu_{\sss\rm R}}
\newcommand\muf{\mu_{\sss\rm F}}

\newcommand\MCatNLO{{\tt MC@NLO}}

\newcommand     \MSB            {\ifmmode {\overline{\rm MS}} \else
                                 $\overline{\rm MS}$  \fi}

\newcommand\POWHEG{{\tt POWHEG}}
\newcommand\POWHEGBOX{{\tt POWHEG BOX}}

\newcommand\MENLOPS{{\tt MENLOPS}}

\newcommand\undecPS{\Phi_{\sss \rm undec. }}
\newcommand\decPS{\Phi_{\sss \rm dec. }}
\newcommand\dectPS{\Phi_{\sss t \to b \bar{\ell} \nu }}
\newcommand\dectbarPS{\Phi_{\sss \bar{t} \to \bar{b} \ell \bar{\nu} }}

\keywords{QCD, Monte Carlo, NLO Computations, Resummation, Collider Physics}

\preprint{
DESY 11-122\\
HU-EP-11/36 \\
LPN11-39 \\
SFB/CPP-11-41 \\
}

\title{Hadronic top-quark pair-production with one jet \\[1ex] 
and parton showering}

\author{Simone Alioli\\
  Deutsches Elektronen-Synchrotron DESY\\
  Platanenallee 6, D-15738 Zeuthen, Germany\\
  E-mail: \email{simone.alioli@desy.de}}

\author{Sven-Olaf Moch\\
  Deutsches Elektronen-Synchrotron DESY\\
  Platanenallee 6, D-15738 Zeuthen, Germany\\
  E-mail: \email{sven-olaf.moch@desy.de}}

\author{Peter Uwer\\
  Institut f\"ur Physik, Humboldt-Universit\"at zu Berlin\\
  Newtonstr. 15, 12489-Berlin Germany\\
  E-mail: \email{Peter.Uwer@physik.hu-berlin.de}}

\abstract{
  We present a calculation of heavy-flavor production in hadronic
  collisions in association with one jet matched to parton shower
  Monte Carlo programs at next-to-leading order in perturbative QCD.
  Top-quark decays are included and spin correlations in the decay
  products are taken into account.  The calculation builds on existing
  results for the radiative corrections to heavy-quark plus one jet
  production and uses the \POWHEGBOX{} for the interface to the parton
  shower programs \PYTHIA{} or \HERWIG{}.  A broad phenomenological
  study for the Large Hadron Collider and the Tevatron is presented.
  In particular we study---as one important sample application---the
  impact of the parton shower on the top-quark charge asymmetry.
}

\begin{document}

\section{Introduction}
\label{sec:intro}

The Large Hadron Collider (LHC) and the Tevatron provide a unique
experimental environment for top-quark physics.  Already with the
currently accumulated luminosity at both colliders, but even more so
with the anticipated high statistics data in the LHC 7~TeV run, the
prospects for precision studies of top-quarks are excellent.  Precise
experimental measurements for top-quark production demand theoretical
predictions with comparable precision.  Within Quantum Chromodynamics
(QCD) this requires the knowledge of the hard scattering process
beyond the leading order (LO) in perturbation theory.  Predictions for
the hadroproduction of top-quark pairs, {\it i.e.}, $pp \to \ttb$, are
available at next-to-leading order (NLO) since
long~\cite{Nason:1989zy,Beenakker:1991ma,Mangano:1991jk,Czakon:2008ii},
and can be improved further thanks to resummation or the use of
approximate next-to-next-to-leading order (NNLO) results, see {\it
  e.g.}, ~\cite{Bonciani:1998vc,Aliev:2010zk}.  The NLO QCD
corrections for the processes $pp \to \ttb
+$jet~\cite{Dittmaier:2007wz,Dittmaier:2008uj} and $pp \to \ttb
+2$~jets~\cite{Bevilacqua:2010ve,Bevilacqua:2011hy} are more recent
achievements.

For the direct comparison with experimental data, however, {\it e.g.},
in order to model experimental acceptances, it is important to provide
also theory predictions for the production of fully exclusive
events. Due to the inherent limitations of the fixed-order approaches,
in the past these were simulated by shower Monte Carlo (SMC) programs,
which were accurate to LO only.  Nowadays, much better theoretical
precision can be reached by merging NLO computations with parton
showers.  The merging method has been pioneered by
\MCatNLO{}~\cite{Frixione:2002ik}.  As an attractive feature the
approach combines the accuracy of exact hard matrix elements for the
large angle scattering including the radiative corrections to first
order in the strong coupling constant $\as$ with the soft and
collinear emission described by the parton shower.  The former
ingredient usually displays a reduced sensitivity with respect to
variations of the renormalization and factorization scales $\mur$ and
$\muf$, while the latter accounts for the correct Sudakov suppression
of collinear and soft emissions. The matching between the regions of
hard and of soft and collinear emissions is smooth.

The follow-up development has seen an extension of the merging
procedure by means of a method, dubbed \POWHEG{}~\cite{Nason:2004rx},
which is independent of the parton SMC generator used.
Moreover, the \POWHEG{} method allows for the generation of positive
weighted events only and thus for a very efficient event generation.
The \POWHEGBOX{}~\cite{Alioli:2010xd} now provides a general framework
for implementing NLO calculations in SMC programs (see
also~\cite{Frederix:2011zi} for progress towards further automation in
\MCatNLO{}).  Processes involving the hadronic production of
top-quarks have been subject to these improvements from very early on.
Top-quark pair-production and single-top production processes have
been implemented in \MCatNLO{}~\cite{Frixione:2003ei,Frixione:2005vw}
and \POWHEG{}~\cite{Frixione:2007nw,Alioli:2009je,Re:2010bp} since
quite some time and are now widely used by experimental
collaborations.

In this article, we are concerned with $\ttb $ pair-production in
association with one hard jet, because a large fraction of inclusive
$\ttb $-events does actually contain one or even more additional jets.
Moreover, due to the larger phase space, the relative importance of
data samples with $\ttb +$jets increases at the LHC with respect to
the Tevatron.  Top-quark pair-production associated with jets is also
an important background to Higgs boson production in vector boson
fusion and for many signals of new physics, {\it e.g.}, those
motivated by supersymmetry.  For the process $\ttb +$jet we are
merging the NLO QCD
corrections~\cite{Dittmaier:2007wz,Dittmaier:2008uj}
(see~\cite{Melnikov:2010iu} for the impact of top-quark decays) with
the parton SMC \HERWIG{}~\cite{Corcella:2000bw} and
\PYTHIA{}~\cite{Sjostrand:1993yb,Sjostrand:2006za} using the
\POWHEGBOX{}.  An independent implementation of this process using the
results of \HELACNLO{}~\cite{Bevilacqua:2010ve,Bevilacqua:2010mx} has
been reported recently in ref.~\cite{Kardos:2011qa}. In this work we
also include spin-correlation effects in the produced events.

The outline is a follows.  In Sec.~\ref{sec:powheg-implementation} we
discuss the implementation details in the \POWHEGBOX{}, which are
specific to the process $\ttb +$jet.  We briefly describe the
validation procedure and carefully examine details of the event
generation in \POWHEG{}.  In Sec.~\ref{sec:results} we present a
phenomenological analysis for the Tevatron with $\sqrt{s}=1.96$~TeV
center-of-mass energy (Sec.~\ref{sec:res-tev}) and the LHC for
$\sqrt{s}=7$~TeV (Sec.~\ref{sec:res-lhc}).  In case of the Tevatron we
investigate the impact of the parton shower on the forward-backward
charge asymmetry which is currently measured at the Tevatron for the
inclusive sample. For the LHC the impact on the charge asymmetry
visible in the rapidity distribution is analyzed. Furthermore we
discuss the inclusion of the top-quark decay and its impact on
differential distributions of the decay products.  Our conclusions
and our outlook on future developments are given in
Sec.~\ref{sec:conclusions}.

\section{Implementation}
\label{sec:powheg-implementation}

The \POWHEGBOX{}~\cite{Alioli:2010xd} provides a well-defined
framework for implementing general NLO calculations in parton SMC
programs.  To that end, it requires as input particular information
about the individual components of the NLO calculation under
consideration, {\it i.e.}, about the Born process, its virtual
radiative corrections and the real emission contributions.
Combination of the latter two and cancellation of the emerging soft
and collinear singularities also requires the definition of a
subtraction scheme.  Within the \POWHEGBOX{} the so-called
Frixione-Kunszt-Signer (FKS) subtraction~\cite{Frixione:1995ms} is the
method of choice.

For the reaction $\ttb +$jet we are building on the computation
of~\cite{Dittmaier:2007wz,Dittmaier:2008uj} for the QCD corrections to
NLO, which provides us with all necessary details.  The list of all
flavor structures of the partonic Born and the real emission processes
can be obtained from generic matrix elements by considering all
possible crossings of light particles into the initial state.  In Born
approximation these are given by
\begin{equation}
\label{eq:lo_processes}
0 \to \ttb g g g
\, , \quad\quad 
0 \to \ttb q \bar q g 
\, ,
\end{equation}
and for the real corrections by
\begin{equation}
\label{eq:nlo_real}
0 \to \ttb g g g g
\, , \quad
0 \to \ttb q \bar q g g
\, , \quad
0 \to \ttb q \bar q q' \bar q'
\, , \quad
0 \to \ttb q \bar q q \bar q
\, ,
\end{equation}
with $q,q'=\{u,d,s,c,b\}$ and $q \ne q'$.

We use {\tt MadGraph}~\cite{Alwall:2007st} for the computation of all
Born squared amplitudes from eq.~(\ref{eq:lo_processes}) and real
emission matrix elements squared from eq.~(\ref{eq:nlo_real}).
Likewise, the color correlated Born squared amplitudes and the
helicity correlated ones, both being necessary ingredients of NLO
calculations employing a subtraction method, have been obtained by
properly modifying {the routines generated by \tt MadGraph}.  Cross
checks have been performed with {\tt
  AutoDipole}~\cite{Hasegawa:2009tx}.  Moreover, factorization in the
soft and collinear limits has been explicitly checked in double and
quadruple precision.

The finite part of the virtual corrections for $\ttb +$jet is
extracted from a {\tt C++} library of
results~\cite{Dittmaier:2007wz,Dittmaier:2008uj}, which is compliant
with the interface to parton SMC programs proposed
in~\cite{Binoth:2010xt}.  In brief, the virtual amplitudes and loop
diagrams associated with eq.~(\ref{eq:lo_processes}) are all generated
analytically and, after manipulation with computer algebra programs,
automatically translated into {\tt C++} code.  The reduction of tensor
integrals up to five-point functions displays good numerical
stability.  Finally, the calculation results in a decomposition of all
loop diagrams according to helicity and color structure times scalar
functions depending on the external momenta only.  The
infrared (IR)-finite scalar integrals are evaluated using the {\tt FF}
package~\cite{vanOldenborgh:1989wn,vanOldenborgh:1990yc} and an
efficient caching system is applied to speed-up the entire
computation.  The current implementation needs approximately 36~ms CPU
time on an {\tt Intel Xeon} processor with 3.00~GHz for each
evaluation of the $\tilde{B}$ function (see eq. 4.5 of ~\cite{Alioli:2010xd})
 at a single phase space point, summing over all contributing
subprocesses.  About half of that time is spent on the computation of
the virtual corrections.

\begin{table}
\begin{center}
\begin{tabular}{cc c c c}
\hline
\hline
 &  $\pt^{\rm gen}$ [GeV] & $\pt^{\rm supp}$ [GeV]  & $\pt^{\rm an}$ [GeV]  &$\sigma^{\rm NLO}$ [pb] \\
\hline
\hline
  & $0$ & $20$  & $20$  & $ 1.793 \pm 0.002$ \\ 
  & $2$ & $0$  & $20$  & $ 1.790 \pm 0.001$ \\  
TEV 1.96 TeV    & $2$ & $20$  & $20$  & $ 1.791 \pm 0.002$ \\  
  & $2$ & $200$  & $20$  & $ 1.793 \pm 0.002$ \\ 
  & $5$ & $0$  & $20$  & $ 1.782 \pm 0.001$ \\ 
  & $5$ & $20$  & $20$  & $ 1.785 \pm 0.001$ \\ 
\hline
\hline
& $0$ & $400$  & $50$ &  $ 52.6 \pm 0.5$ \\
 & $5$ & $400$  & $50$ &  $ 52.7 \pm 0.5$ \\
LHC 7 TeV & $5$ & $100$  & $50$ &  $ 53.1 \pm 0.2$ \\ 
 & $10$ & $0$  & $50$ &  $ 52.9 \pm 0.4$ \\
 & $10$ & $400$  & $50$ &  $ 52.5 \pm 0.1$ \\
 & $15$ & $0$  & $50$ &  $ 52.6 \pm 0.4$ \\
\hline
\hline
       & $0$ &$400$   &$50$ & $ 379.8 \pm 1.6$ \\
LHC 14 TeV & $5$ &$100$   &$50$ & $ 376.1 \pm 0.2$ \\
 & $5$ &$400$   &$50$ & $ 377.2 \pm 1.6$ \\
\hline
\hline
\end{tabular}
\end{center}

\caption{
\label{tab:gencut}
  The dependence of the NLO cross section at Tevatron and LHC on the
  generation cut $\pt^{\rm gen}$ and on the Born suppression factor
  $\pt^{\rm supp}$ for $\mur=\muf=m_t=174~$GeV and CTEQ6M PDFs. Jets
  are reconstructed by the inclusive $\kt$-algorithm with $R=1$.  }
\end{table}
With the input described so far, our \POWHEGBOX{} implementation can
be used to generate hadronic events accurate to NLO.  There is,
however, an important technical issue related to multi-leg processes
which are interfaced to a parton shower: 
Reactions such as $\ttb+$parton possess soft and collinear divergences already at LO.  Since
the \POWHEG{} method relies on generating all events starting from a
momentum configuration of the underlying Born process, a generation
cut, $\pt^{\rm gen}$, needs to be placed on the transverse momentum of
the final-state partons, a procedure well-known from standard SMC
generators, see {\it e.g.}, the discussion in~\cite{Alioli:2010qp}.

This cut for the event generation, which in practice takes values 
{\it e.g.}, $\pt^{\rm gen} \approx 1~$GeV, is unphysical and has to be
always much smaller than the analysis cut on the transverse momentum,
$\pt^{\rm an}$, employed in the definition of the additional jet
associated with the $\ttb$-pair.  The condition $\pt^{\rm gen}~
\le~\pt^{\rm an}$ is necessary though not sufficient to provide a
suitable sample of unweighted events, because the parton shower can in
fact increase the transverse momentum in an event generated at Born
level with a given $\pt \le \pt^{\rm gen}$. This may result in a
different number of events that would have passed the analysis cut if
the generation cut were different.  Therefore, it is essential to show
that the actual dependence on $\pt^{\rm gen}$ is negligible for a
given fixed analysis cut $\pt^{\rm an}$.  We have carefully
investigated this issue, reporting some examples in 
Tab.~\ref{tab:gencut}, which shows the independence of the cross
section on the generation cut $\pt^{\rm gen}$ for reasonable values of
the cut. In particular, this is satisfied for $\pt^{\rm gen}$ smaller than
$\pt^{\rm an}/4$. Tab.~\ref{tab:gencut} provides also a strong cross check of
existing results in~\cite{Kardos:2011qa}.

Also a second option for dealing with processes with soft and
collinear divergences at Born level is available by generating
weighted rather than unweighted events and, thereby suppressing the
divergences.  The \POWHEGBOX{} uses a suppressed cross section ${\bar
  B}_{\rm supp}$ for the generation of the underlying Born
configurations~\cite{Alioli:2010qp},
\begin{equation}
  \label{eq:Bornsupp}
  {\bar B}_{\rm supp} = {\bar B} \times F(\pt)
\, ,
\end{equation}
where $\pt$ is a measure of the hardness of the extra emission, ${\bar
  B}$ denotes the inclusive NLO cross section at fixed underlying Born
variables (see eq.~4.2 of ~\cite{Alioli:2010xd}), and $F(\pt)$ has to
be chosen such that it ensures finiteness of ${\bar B}_{\rm supp}$ in
the limit $\pt \to 0$.  Thus, ${\bar B}_{\rm supp}$ becomes integrable
and, of course, the generated events need to be weighted by
$1/F(\pt)$.  The \POWHEGBOX{} implementation foresees the choice
\begin{equation}
  \label{eq:Fsupp}
  F(\pt) = \left( \pt^2  \over \pt^2 + \(\pt^{\rm supp}\)^2 \right)^n
  \, ,  
\end{equation}
where the variable $\pt^{\rm supp}$ controls the extent of the
suppression\footnote{ In the \POWHEGBOX{} the value of $\pt^{\rm
    supp}$ can be assigned at run time trough the $\ptsupp$ entry of
  the input card.}  and $n$ is process specific, $n=2$ for $\ttb
+$jet.  In our event generation we have also applied the Born
suppression factor of eq.~(\ref{eq:Fsupp}) and checked its
implementation for different $\pt^{\rm supp}$ values with very good
numerical accuracy, as shown in Tab.~\ref{tab:gencut}.

\subsection{Checks}

The current implementation has been exposed to a large number of checks.
These include, first of all, comparisons for a sizable number of
differential distributions to NLO with the available fixed order
results.  We have used the settings
of~\cite{Dittmaier:2007wz,Dittmaier:2008uj}, {\it i.e.}, the
renormalization and factorization scales $\mu_R = \mu_F = m_t =
174~$GeV, the analysis cut of $\pt > 20~$GeV and $\pt > 50~$GeV for
the Tevatron and LHC configuration, respectively, the parton
distribution function (PDF) set CTEQ6M~\cite{Pumplin:2002vw}, the jet
algorithm of~\cite{Ellis:1993tq} (inclusive-$k_T$) with $R = 1$ and we
have assumed here the top-quarks to be always tagged and thus excluded
from jet reconstruction.  Perfect agreement
with~\cite{Dittmaier:2007wz,Dittmaier:2008uj} has been found typically
at the per-mille level.  Also note that the computation
of~\cite{Dittmaier:2007wz,Dittmaier:2008uj} has employed dipole
subtraction with massive partons~\cite{Catani:2002hc} for the
cancellation of the soft and collinear divergences between the real
and virtual contributions.  As mentioned, the \POWHEGBOX{} relies on
FKS subtraction~\cite{Frixione:1995ms} and, of course, the results for
physical cross sections at NLO must be independent of the chosen
scheme, which constitutes another strong check of the current
implementation.

The other aspect which demands careful checking is the parton shower.
It adds all-order perturbative corrections, though to leading logarithmic
accuracy only, and the \POWHEG{} approach offers one particular choice
of how these corrections beyond NLO are included.  Since the merging
with parton showers modifies the QCD predictions, sometimes even for
inclusive quantities, one should check during the validation process
that these modifications reflect real physics effects and moreover,
are compatible with (un)known higher-order corrections.  To that end,
we have also compared the NLO fixed order and the \POWHEG{}
predictions after the first emission, {\it i.e.}, at the level they
are written to the Les Houches event file~\cite{Alwall:2006yp} (LHEF
from now on), for suitable distributions in order to assess
similarities and differences between the results.  Of particular
interest in this procedure are those observables, which display great
sensitivity to the parton shower effects.  Some results will be shown
below in Sec.~\ref{sec:results}.

\section{Results}
\label{sec:results}

Throughout this section we present our findings.  In comparing to the
fixed order NLO results we are using the parton shower programs
\HERWIG{}~\cite{Corcella:2000bw} (version 6.5.20) and
\PYTHIA{}~\cite{Sjostrand:1993yb,Sjostrand:2006za} (version 6.4.25),
with the default setting of parameters. Moreover, in order to
highlight the shower effects and to disentangle them from the
underlying event (UE) or multiple particle interactions (MPI), we have
generally switched off these features in the SMC programs.  Of course
MPI and UE, together with the usage of the appropriate SMC tuning,
will play an important r\^ole when comparing \POWHEG{} predictions
with data. For the kind of study we are interested in, we have decided
not to include them in the following plots and we have concentrated on
presenting distributions which are not extremely sensitive to these
effects, where possible. The one exception here is the discussion of
the charge asymmetries, in Secs.~\ref{sec:afbtev} and~\ref{sec:aclhc}.
There, we will also consider the predictions obtained including also
MPI and UE in our comparisons in order to asses the stability of our
results with respect to these effects.

We have produced different samples of events, to study both, the
Tevatron and LHC collider configurations.  Each sample contains 20M
positive and negative weighted events, produced without folding (see
Appendix~\ref{app:folding}) and with a generation cut $\pt^{\rm
  gen}=2$~GeV and $5$~GeV, respectively, for the Tevatron and the LHC.
As already mentioned, the jet reconstruction cut in the analysis has
been assumed to be $\pt > 20$~GeV and $50$~GeV.  We also make use of
the $\ptsupp$ option of the \POWHEGBOX{} to damp the low transverse
momentum regions in the cross section, artificially increasing the
probability of a hard jet.  For this reason, the events have a weight
which is the inverse of eq.~(\ref{eq:Fsupp}).  The choices $\pt^{\rm
  supp}=20 (100)$~GeV for the Tevatron (LHC) were adopted.

For the sake of comparisons and, in order to keep the analysis routine
rather insensitive to possible contamination in the jet reconstruction
procedure, we always force the semi-leptonic decay of the
(anti-)top-quark when interfacing to a SMC program in the following.
Also, as already done for the NLO comparisons of
Sec.~\ref{sec:powheg-implementation}, we will always use the jet
algorithm of~\cite{Ellis:1993tq} (inclusive-$k_T$) with $R = 1$ and
the $E_T$-recombination scheme.\footnote{Comparisons have been done
  with other jet clustering algorithms, {\it e.g.}, the anti-$k_T$ jet
  algorithm~\cite{Cacciari:2008gp} with $p$-scheme recombination, for
  different values of the parameter $R$, without significantly
  changing the general conclusions drawn here.}  In addition, after
interfacing to a SMC program and performing the hadronization stage,
we always exclude from the jet list those jets which happen to contain
a $b$-flavored hadron, whose origin can be traced back to the
(anti-)top-quark decay.\footnote{This exclusion is applied also when
  the top-quark decay is performed by \POWHEG{} itself and
  bottom-quarks coming from top-quarks can be present already at the
  LHEF stage, as it will be the case in Sec~\ref{sec:topdecay}. Other
  bottom-quarks present at partonic level (or B hadron not originating
  from a (anti-)top-quark) are instead always consistently considered
  as (originating from) light partons and therefore included in the
  light jets count.}.

As we focus entirely on the effect of the parton SMC and the NLO
merging, our phenomenological analysis in Secs.~\ref{sec:res-tev}
and~\ref{sec:res-lhc}
follows~\cite{Dittmaier:2007wz,Dittmaier:2008uj}, fixes the
renormalization and factorization scales $\mu_R = \mu_F = m_t =
174~$GeV, and uses the PDF set CTEQ6M~\cite{Pumplin:2002vw}.  Studies
of the theoretical uncertainty of the cross section for $pp \to \ttb
+$ jet due to corrections of higher orders (beyond NLO in QCD) have
already been conducted in~\cite{Dittmaier:2007wz,Dittmaier:2008uj}
with standard  means, {\it i.e.}, determining the
variations for the scale choices of $\mur=\muf=m_t/2$ and
$\mur=\muf=2m_t$.  In Tab.~\ref{tab:scalevar} we report results for
the Tevatron $1.96$~TeV as well as for the $7$~TeV and $14$~TeV LHC
configurations allowing for the independent variations of $\mur$ and
$\muf$ of factors $m_t/2$ and $2m_t$ and, at the same time, excluding
relative ratios of $\mur/\muf$ larger than $2$ or smaller than
$1/2$. We find a remarkable small variation, typically
  below 10\% .  On the basis of the scale variations uncertainty one
  can thus conclude that higher order effects should be small. 

\begin{table}
\begin{center}
\begin{tabular}{ccc}
\hline
\hline
& $\pt^{\rm an}\rm~[GeV]$& $\sigma_{\sss \rm  NLO} \rm~[pb]$ \\
\hline
\hline
TEV $1.96$ TeV & $20$ &$1.791(2) ^{+0.160}_{-0.318}$ \\
LHC $7$ TeV &  $50$ &  $53.1(2) ^{+4.1}_{-8.9}$ \\
LHC $14$ TeV & $50$ &  $376.1(2) ^{+20.1}_{-45.4}$\\
 \hline
\hline
\end{tabular}
\end{center}

\caption{
\label{tab:scalevar}
  The dependence of the NLO cross section at Tevatron and LHC on the
  renormalization $\mur$ and factorization scales $\muf$.  The
  envelope has been constructed by considering independent variations
  of $\mur$ and $\muf$ around $m_t=174~$GeV, by a factor of two in both
  directions. Combinations resulting in ratios of  
  $\mur/\muf$ larger than $2$ or smaller than $1/2$ have been excluded.  
  Jets are reconstructed by the inclusive $\kt$-algorithm with
  $R=1$ and 
  CTEQ6M
  PDFs are adopted.  }
\end{table}

The other main source of theoretical uncertainties,
related to the non-perturbative
parameters such as the PDFs and the associated value of the strong
coupling constant $\as(M_Z)$, have repeatedly been addressed in the
literature (see {\it e.g.},~\cite{Aliev:2010zk}) using modern PDF sets
available to
NNLO~\cite{Alekhin:2009ni,JimenezDelgado:2008hf,Martin:2009iq}.  For a
more recent study concerning the combined uncertainty due to
propagated PDF uncertainties and uncertainties in $\as(M_Z)$ we refer
the interested reader to the results discussed in \cite{Watt:2011kp}.
The top-quark mass $m_t$ is always taken as a pole mass, and we refer
to~\cite{Langenfeld:2009wd} for a discussion of the running mass in
the \MSB-scheme.

\subsection{Tevatron}
\label{sec:res-tev}

We will begin our studies with the Tevatron ($\sqrt{s}=1.96$~TeV),
which has provided us already with several measurements of
differential distributions.  These are, in detail, the first
measurement of the $\ttb$ cross section $d \sigma/d m_{\ttb}$,
differential in $\ttb$-pair invariant mass $m_{\ttb}$,
see~\cite{Aaltonen:2009iz}, the dependence of the $\ttb$ production
cross section on the transverse momentum of the
top-quark~\cite{Abazov:2010js} and most prominently, the measurements
of the forward-backward charge asymmetry in top-quark
pair-production~\cite{Abazov:2007qb,Aaltonen:2008hc} along with its
dependence on the $\ttb$-pair invariant mass
$m_{\ttb}$~\cite{Aaltonen:2011kc,Abazov:2011rq}.

In all the plots throughout this article no acceptance cuts are
imposed, other than those necessary to define the hard jet.  The lines
appearing the legenda are labeled according their origin:
\tmtexttt{NLO} for the fixed order computation, \tmtexttt{LHEF} for the
results after the \POWHEG{} hardest emission without any showering and
\tmtexttt{PWG+HER}, \tmtexttt{PWG+PYT} for the results obtained by
generating the hardest emission with the \POWHEG{} method and then
performing the remaining shower and hadronization stages with the
\HERWIG{} or \PYTHIA{} programs, respectively. The lower inset in each
plot shows the relative difference with respect to the first entry in
the legend, which is usually depicted as a solid black line in the
upper inset.  The black dots in the lower inset stand instead as a
reference for the zero line. The Monte Carlo integration errors are
shown as vertical bars.

\begin{figure}[htb]
\centering
\vspace*{10mm}
    {
    \includegraphics[width=\figwidth]{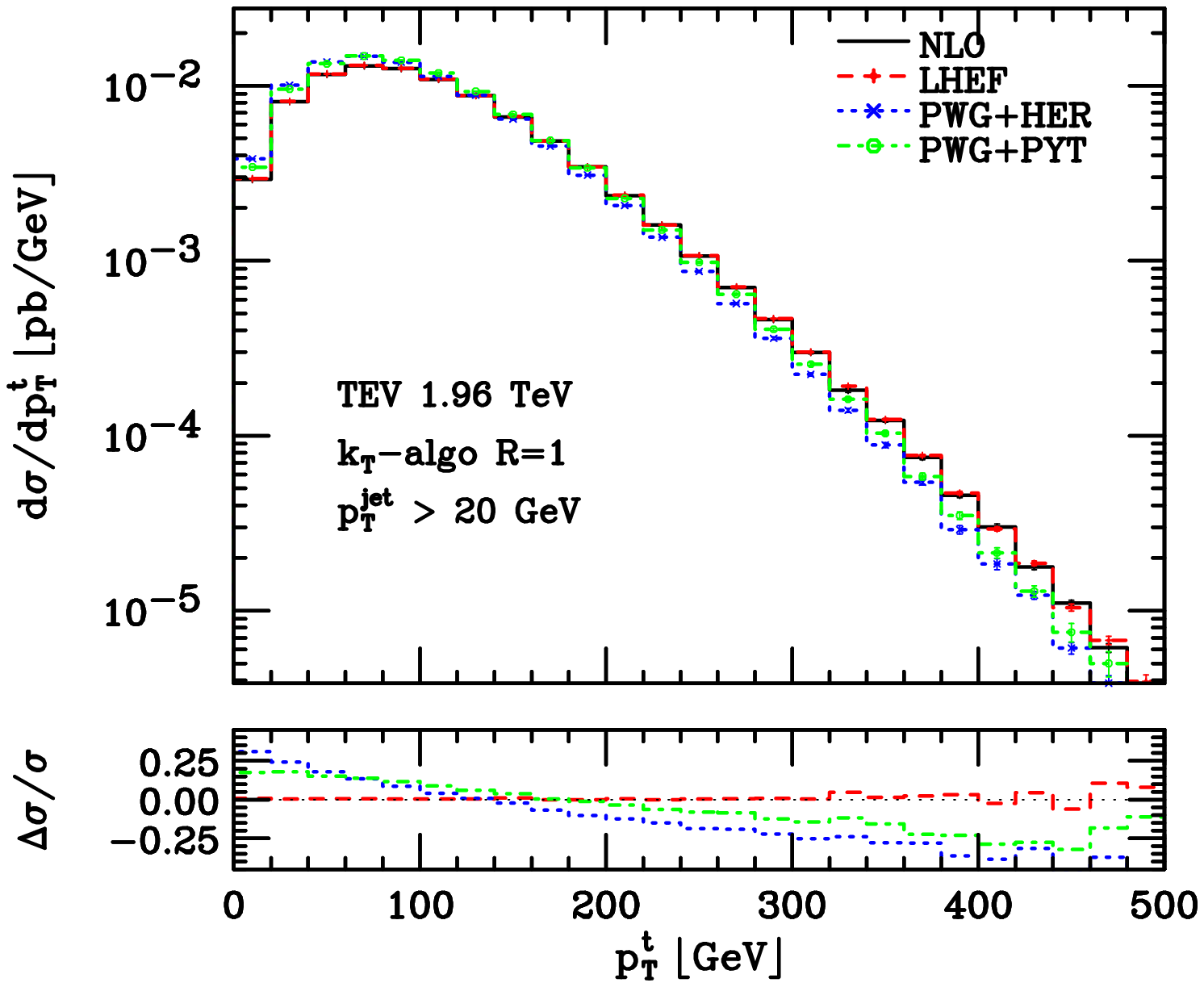}
    \hfill
    \includegraphics[width=\figwidth]{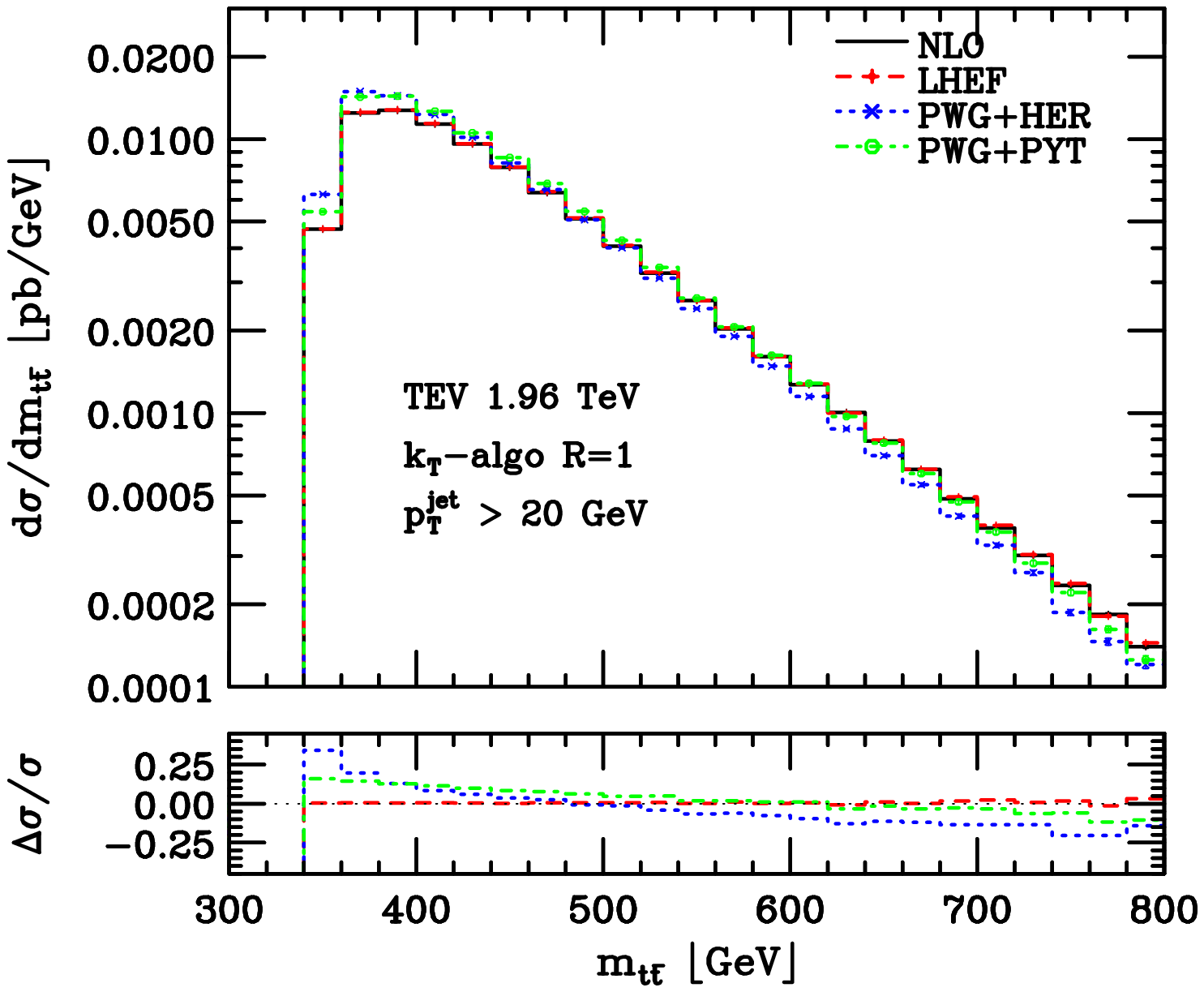}
    }
    \caption{ \small
      \label{fig:TEV-pt_t-y_t}
The differential cross sections as function of the transverse momentum $\pt^{\;t}$ (\figleft{} panel) and of the $\ttb$-pair
invariant mass $m_{\ttb}$ (\figright{} panel),  at the 
Tevatron ($\sqrt{s}=1.96$~TeV).
    }
\end{figure}
In Fig.~\ref{fig:TEV-pt_t-y_t}, we plot the differential cross section
as a function of the transverse momentum $\pt^{\;t}$ of the top-quark
(\figleft{} panel) and of the the $\ttb$-pair invariant mass
$m_{\ttb}$ (\figright{} panel).  A first comment can be made upon the
expected good agreement between the NLO results and the LHEF ones for
such inclusive quantities. The effect of the shower does not change
considerably these results. Only at the very low and high ends we
observe a significant change which is however still below 30\%.

\begin{figure}[htb]
\centering
\vspace*{10mm}
    {
    \includegraphics[width=\figwidth]{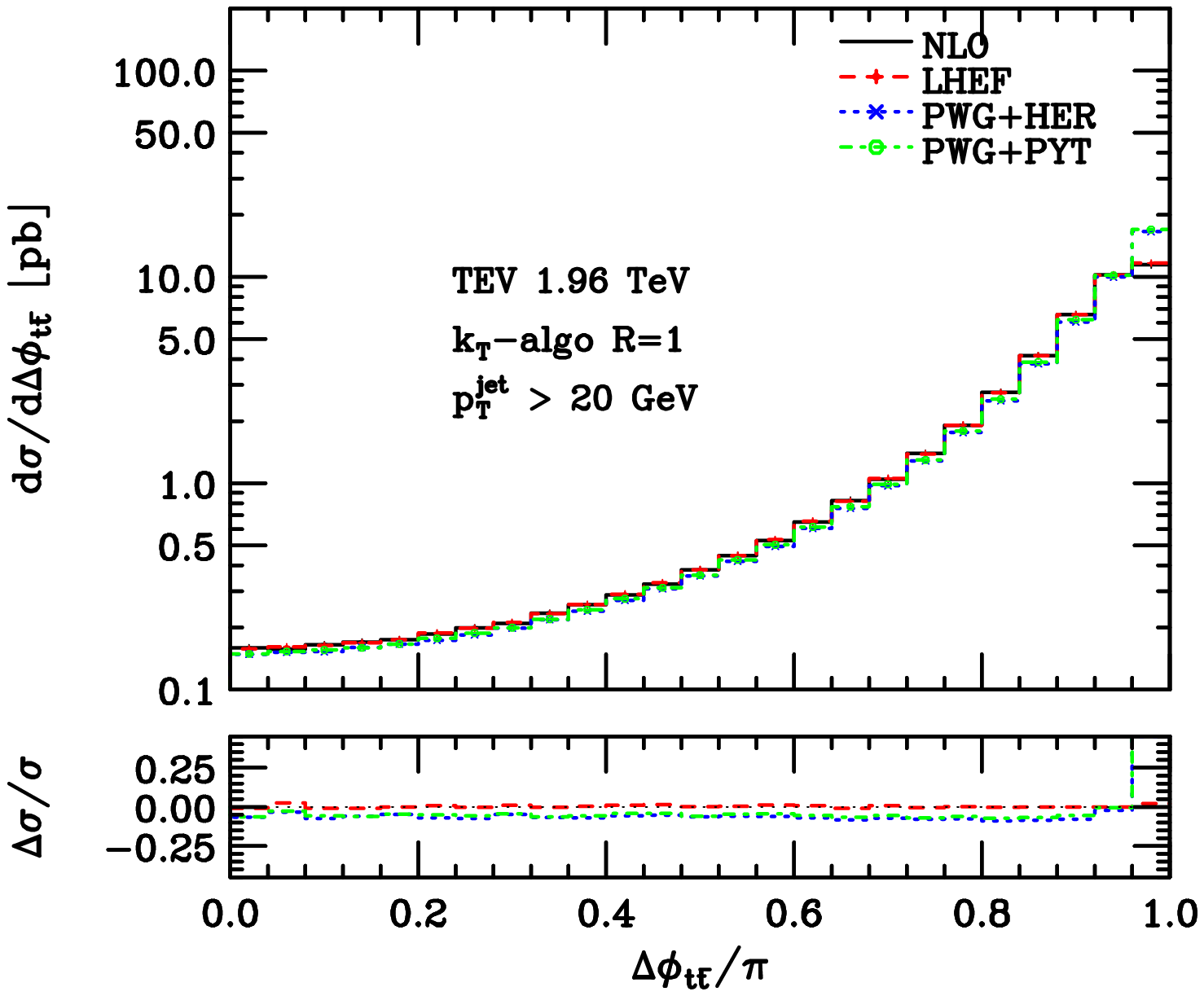}
    \hfill
    \includegraphics[width=\figwidth]{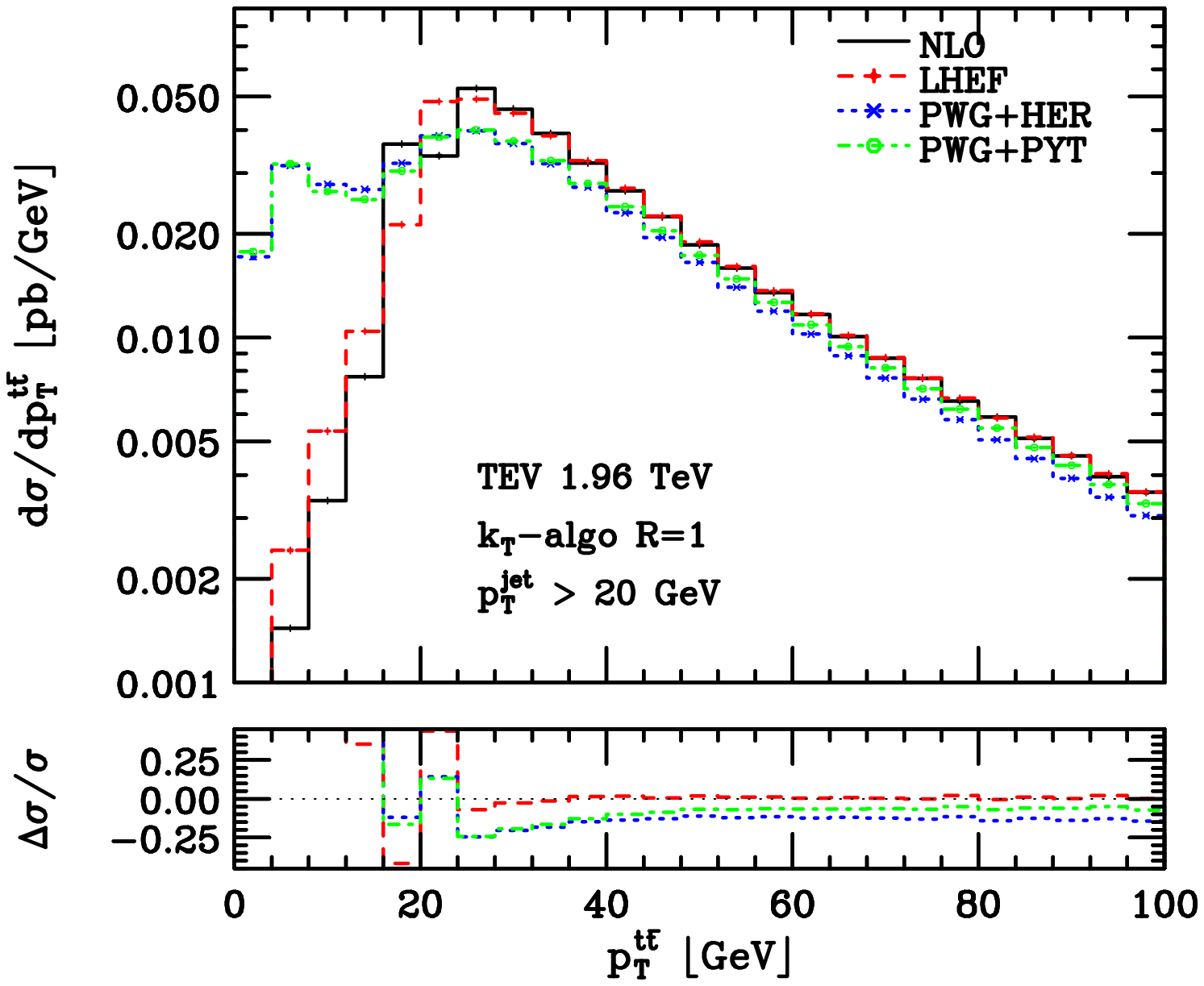}
    }
    \caption{ \small
      \label{fig:TEV-Dphitt-pttt}
The differential cross sections as function of the $\ttb$-pair
azimuthal separation $\Delta\phi_{t
  \bar t}$  and 
transverse momentum $\pt^{\ttb}$, at the Tevatron
($\sqrt{s}=1.96$~TeV).  }
\end{figure}

Similar conclusions can also be drawn looking at the \figleft{} panel
of Fig.~\ref{fig:TEV-Dphitt-pttt}, where we plot the differential
cross sections $d\sigma/d\Delta\phi_{t \bar t} $ as a function of the
azimuthal separation between the top- and anti-top-quarks.  In
particular, this last observable is very stable with respect to the
inclusion of the parton shower. With exception of the last bin, we
observe corrections of a few percent only. The understanding is that
the parton shower is not able to produce a significant change of the
relative directions of the two outgoing heavy quarks.  In the
\figright{} panel instead, we show the differential cross sections
$d\sigma/d\pt^{\,\ttb } $ as a function of the transverse momentum of
the $\ttb$-pair. Shower effects at low $\pt^{\,\ttb }$ values are now
clearly visible.  Furthermore, by inspecting the lower inset in the
plot, showing the relative difference with NLO predictions, it is also
possible to see a hint of the instabilities that may arise in the
fixed order calculation around the $\pt = 20$~GeV jet cut. 
\clearpage This
behaviour is easily explained considering that the region below the
cut may be populated only when more than one hard jet is resolved,
thus allowing for an imbalance between the hardest jet and the
$\ttb$-pair. Indeed, for the $3$-partons $\ttb j$ configuration,
$\pt^\ttb=\pt^{j1}$ by momentum conservation.  This effectively makes
the $d\sigma/d\pt^{\,\ttb } $ distribution for $\pt^{\,\ttb }$ lower
than the jet $\pt$ cut a LO quantity, starting at ${\cal O}\(\as^4\)$.
We have verified that the instability gets worse by reducing the bin
size and that neither the LHEF nor the showered results show a similar
behaviour.  The same feature will also be present in LHC predictions,
around the $\pt = 50$~GeV jet cut, as we will show in the \figright{}
panel of Fig.~\ref{fig:LHC-mtt-pttt}.  Such instabilities are well
known to arise at any fixed order of the perturbative expansion
whenever the observable under consideration has a non-smooth behaviour
inside the physical region or whenever the phase space boundary for a
certain number of partons lies inside that for a larger
number~\cite{Catani:1997xc}.  In our case the LO $d\sigma/\pt^\ttb$
distribution is discontinuous at the jet $\pt^{\rm cut}$ boundary,
since the phase space for $3$ partons does not allow $\pt^{\,\ttb } =
\pt^{j1} < \pt^{\rm cut}$, while it is different from zero at
$\pt^{\,\ttb} \ge \pt^{\rm cut}$.  It should be stressed that despite
  these effects are due to unbalanced cancellation of higher order
  soft/collinear divergences at the critical point(s), they are
  integrable in any finite neighborhood.  In ref.~\cite{Catani:1997xc}
  it was also demonstrated that the all-order resummation of soft
  divergences restores the infrared finiteness of the predictions also
  at the critical point by the appearance of continuous and infinitely
  differentiable structures, that were dubbed Sudakov shoulders.  We
  have indeed verified that the resummation implicit in the \POWHEG{}
  approach does reproduce these structures, as can be observed in the
  figures.

\begin{figure}[htb]
\centering
\vspace*{10mm}
    {
    \includegraphics[width=\figwidth]{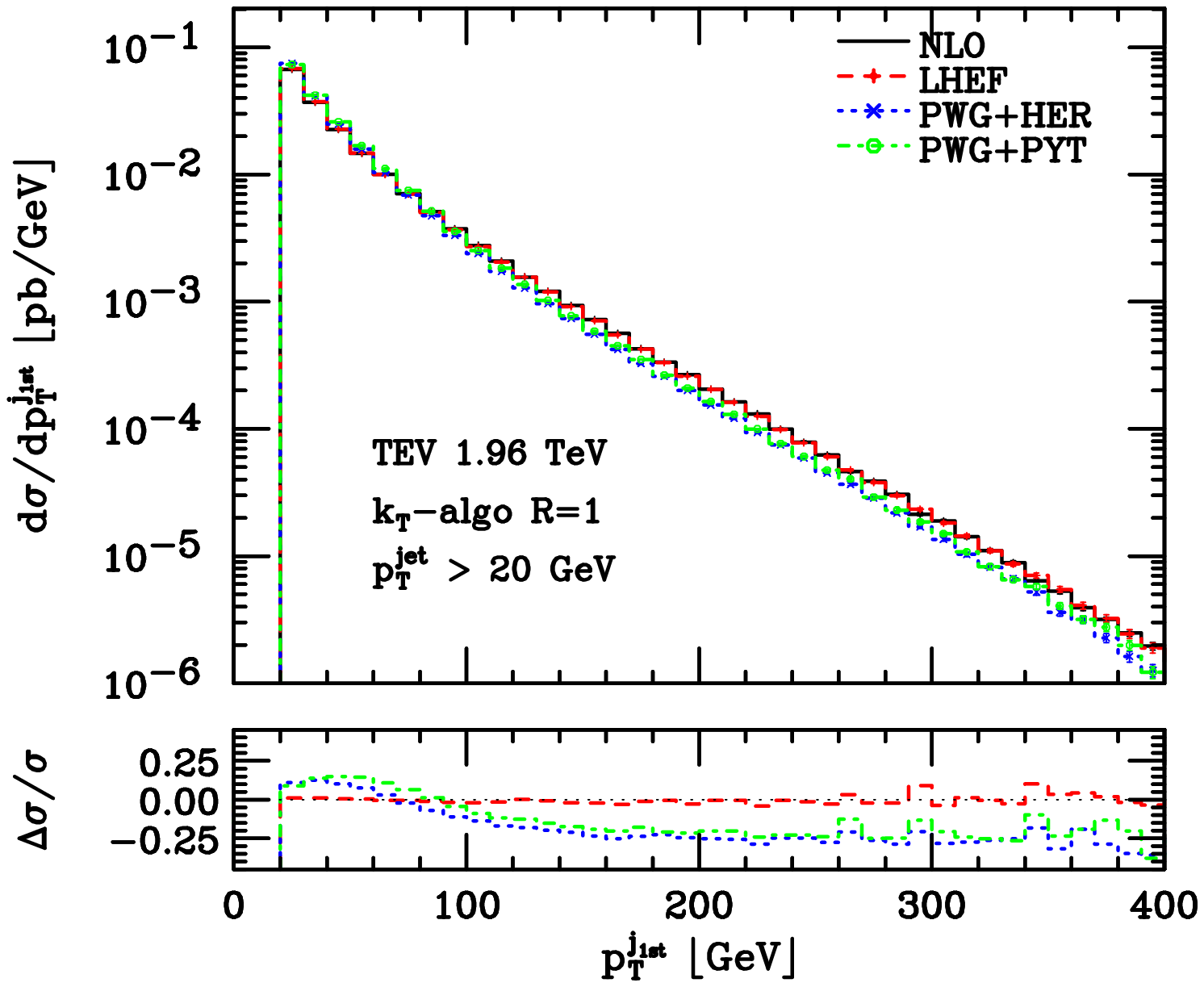}
    \hfill
    \includegraphics[width=\figwidth]{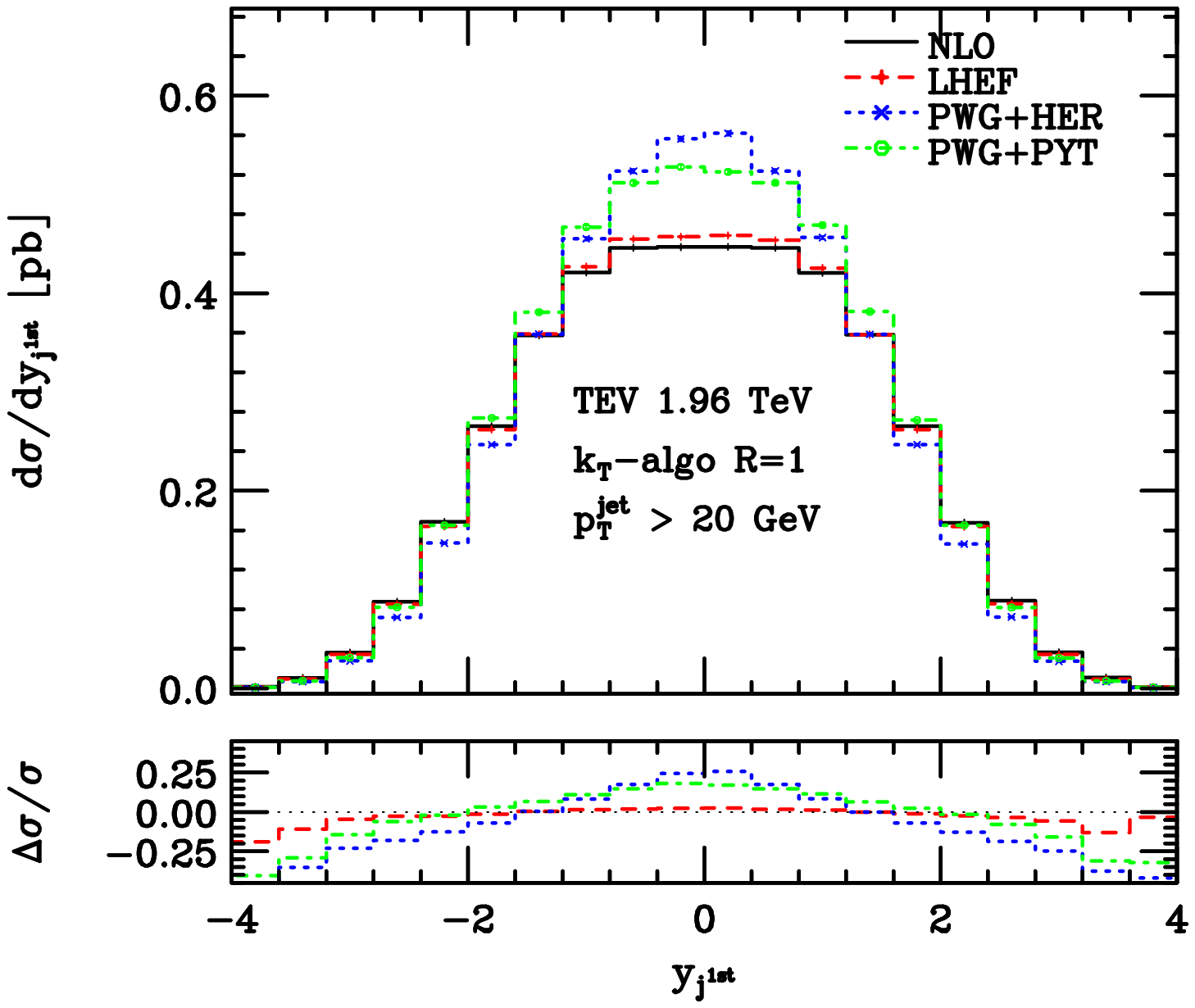}
    }
    \caption{ \small
      \label{fig:TEV-j1_pt-j1_y}
The differential cross sections as function of the hardest jet
transverse momentum and rapidity at the Tevatron ($\sqrt{s}=1.96$~TeV).  }
\end{figure}

From Fig.~\ref{fig:TEV-j1_pt-j1_y} on we investigate the jet structure
of the process.  To that end, in Fig.~\ref{fig:TEV-j1_pt-j1_y}, we
display the differential cross sections $d\sigma/d\pt^{j_{1\rm st}} $
(\figleft{} panel) and $d\sigma/dy_{j^{1\rm st}} $ (\figright{} panel) as function
of the transverse momentum and rapidity of the hardest jet.  Here we
notice a moderate shape distortion and a different normalization in
going from parton level to hadronic events.
Both these effects can be
understood by considering that the jet clustering procedure is
significantly affected when going from a low multiplicity sample to a fully hadronized event. 
This is also reflected in the number of resolved
jets above the minimum $\pt$ cut and in their kinematics.

\begin{figure}[htb]
\centering
\vspace*{10mm}
    {
    \includegraphics[width=\figwidth]{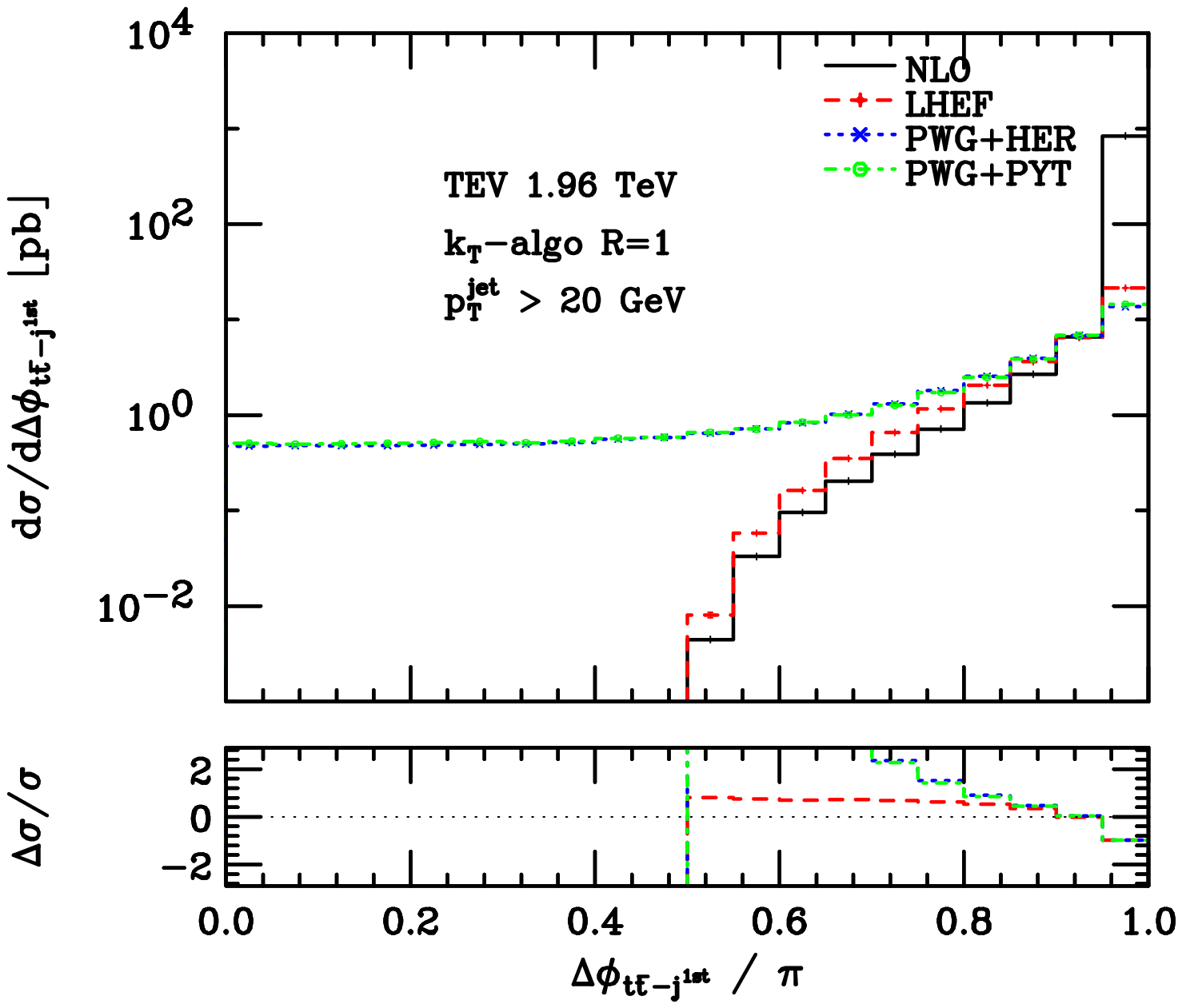}
    \hfill
    \includegraphics[width=\figwidth]{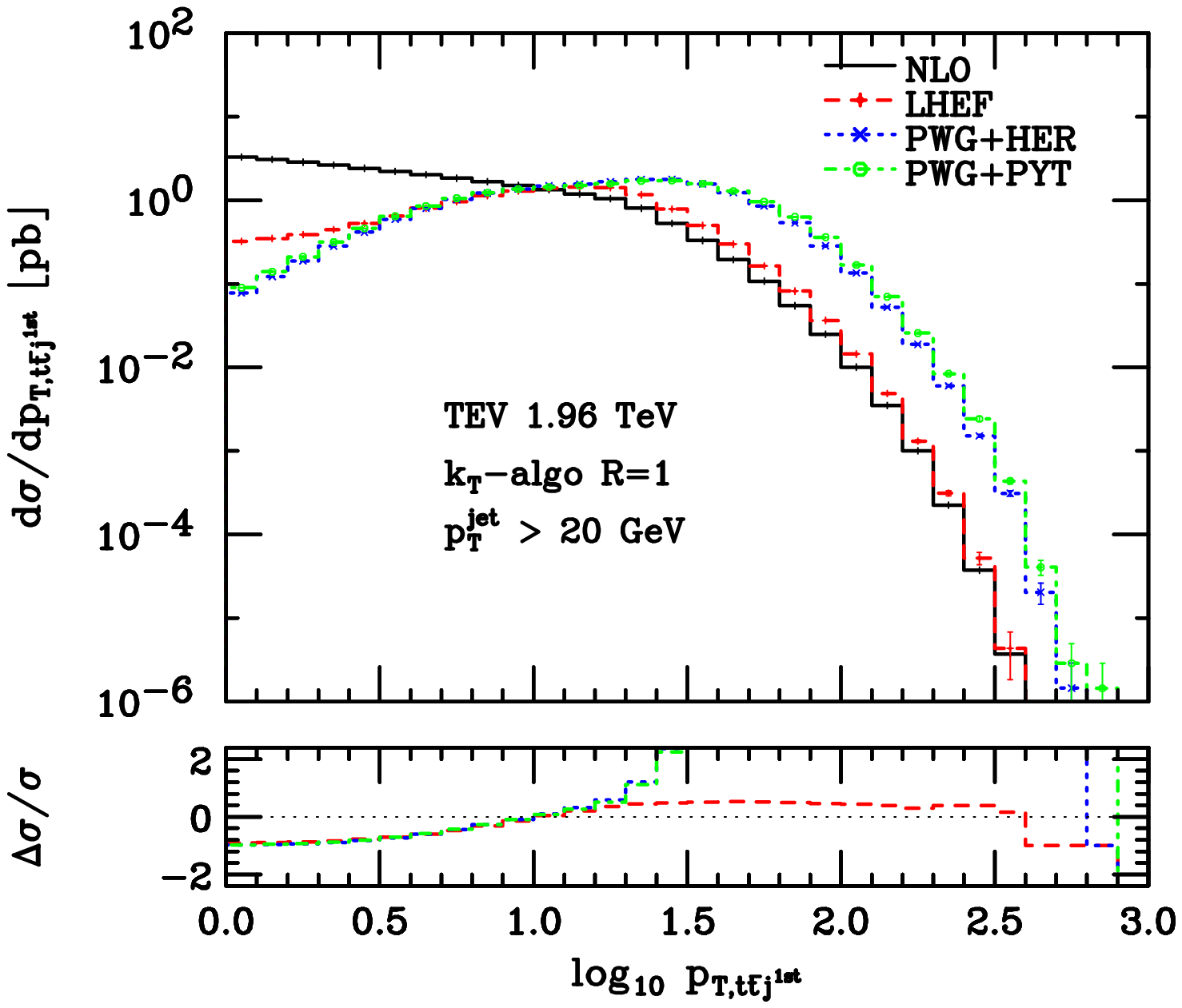}
    }
    \caption{ \small
      \label{fig:TEV-ttbarj1_Dphi-ttbarj1_logpt}
The differential cross sections as function of the azimuthal distance
of the hardest jet from the $\ttb$-pair (\figleft{} panel) and as a
function of the transverse momentum of the system made by $\ttb$-pair
and the hardest jet (\figright{} panel) at the Tevatron
($\sqrt{s}=1.96$~TeV). }
\end{figure}

In the \figleft{} panel of
Fig.~\ref{fig:TEV-ttbarj1_Dphi-ttbarj1_logpt} we show the azimuthal
difference $\Delta\phi_{\ttb - j^{1\rm st}}$ between the hardest jet
and the $\ttb $ system. $\Delta\phi_{\ttb - j^{1\rm st}}$ is very
sensitive to shower effects, since at partonic level the hardest jet
is constrained to reside in the opposite hemisphere w.r.t. the
$\ttb$-pair momentum. In the \figright{} panel we plot the
differential cross section $d\sigma/d \pt^{\ttb j_{1\rm st}}$ as
function of the transverse momentum of the system composed of the
top-quark, the anti-top-quark and the hardest jet.  This observable is
of particular importance, as it is strongly correlated with the
transverse momentum of the next-to-hardest jet.  The latter, in turn,
is the quantity where the Sudakov suppression effects should be most
evident.  Due to the jet reconstruction procedure, however, it turns
out that those effects are best seen in the low $\pt^{\ttb j_{1\rm
    st}}$ region.  The disagreement in the high $\pt^{\ttb j_{1\rm
    st}}$ tail between NLO and LHEF results, on the one side, and
showered results, on the other, is instead due to showering and
hadronization effects and is partly related to our choice of excluding
the top-quark from jet reconstruction.  A similar effect has also been
observed (and explained) in single-top production (see Sec.~4
of~\cite{Alioli:2010xa}) and we have verified that the same
explanation can readily be applied here too, for the ${\ttb j_{1\rm
    st}}$ system: At high $\pt^{\ttb j_{1\rm st}}$ there exists an
artificial imbalance due to many hadrons coming from the hardest
parton which are not clustered together with the hardest jet.  This
creates an effective $\pt$ for the system containing the hardest jet,
giving rise to the harder $\pt^{\ttb j_{1\rm st}}$ tail observed.

\begin{figure}[htb]
\centering
\vspace*{10mm}
    {
    \includegraphics[width=\figwidth]{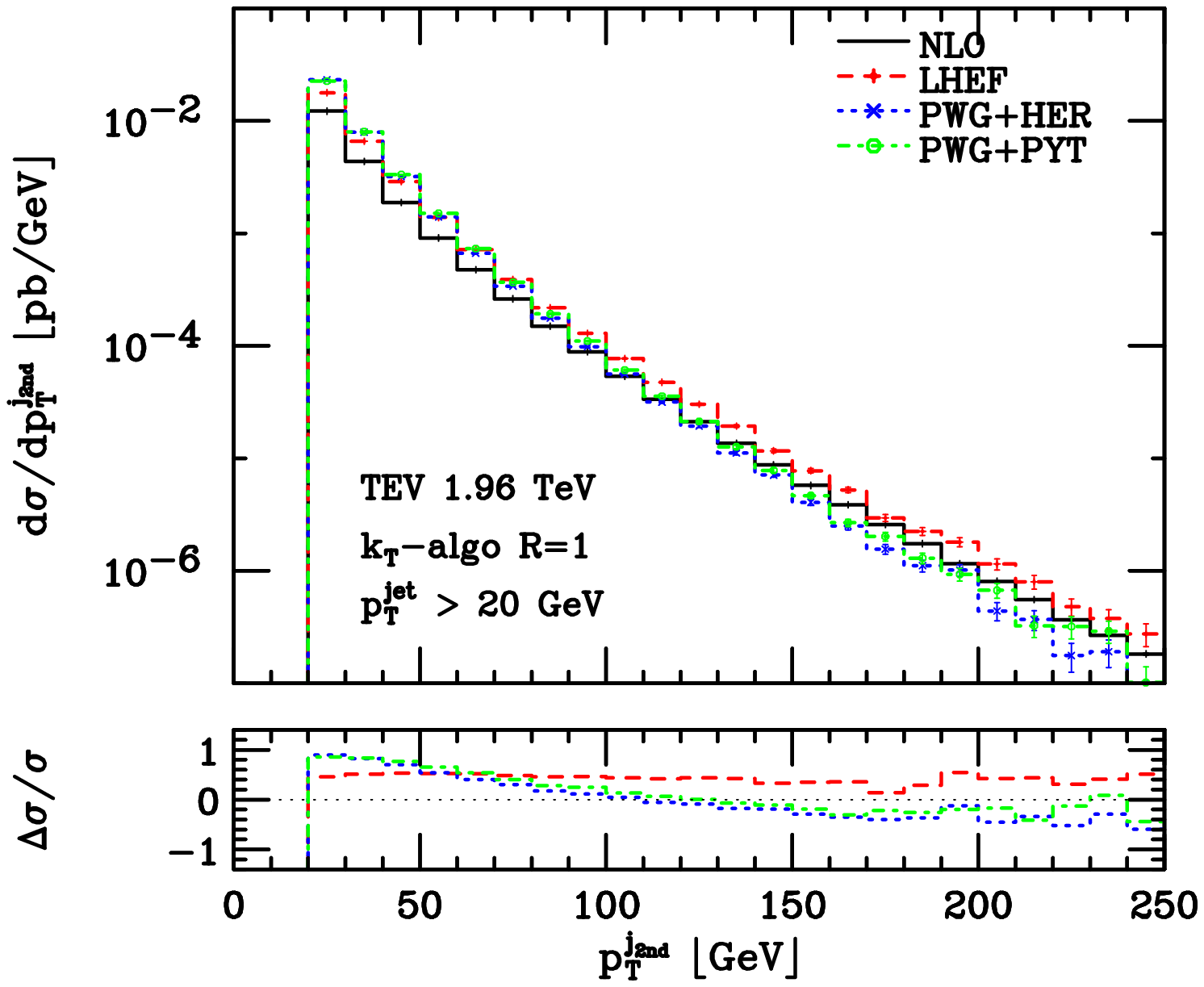}
    \hfill
    \includegraphics[width=\figwidth]{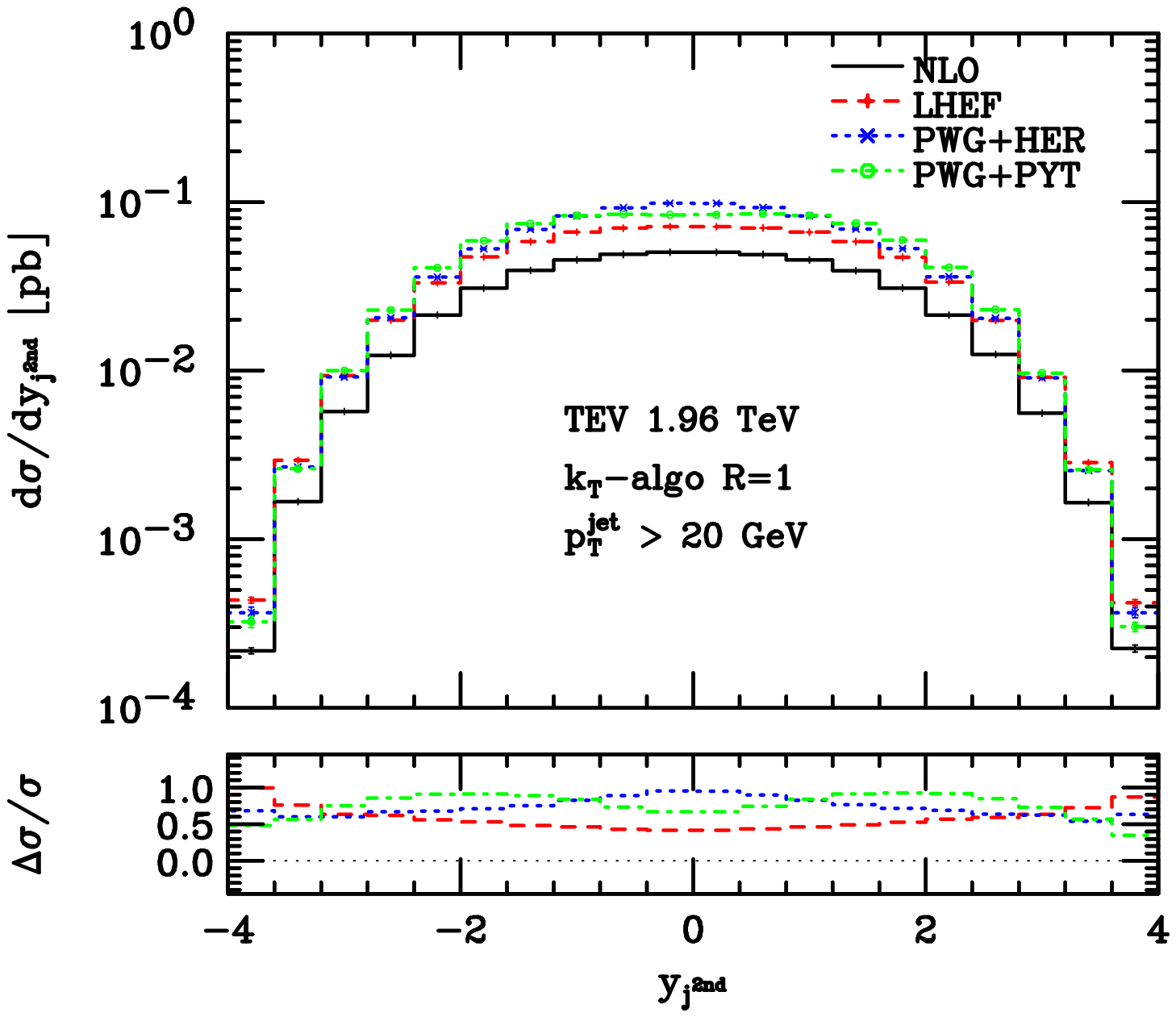}
    }
    \caption{ \small
      \label{fig:TEV-j2_pt-j2_y}
      The differential cross sections as function of the
      next-to-hardest jet transverse momentum and rapidity at the
      Tevatron ($\sqrt{s}=1.96$~TeV).  }
\end{figure}
\clearpage
Finally, in Fig.~\ref{fig:TEV-j2_pt-j2_y} we plot the differential
cross section as function of the transverse momentum $\pt^{j_{2\rm
    nd}}$ and rapidity $y_{j^{2\rm nd}}$ of the next-to-hardest jet.
No high-$\pt$ enhancement, or suppression, is observed in this
case. However, large differences in the normalization of distributions
are present when going from NLO results to LHEF and showered ones. This
can again be explained with the different number of configurations
with are resolved by the reconstruction algorithm as containing two hard jets.

\begin{sidewaystable}[htb]
\begin{center}
\begin{tabular}{ccccccc}
\hline
\hline
Tevatron $1.96$~TeV &NLO [\%]&LHEF [\%]&PWG+HER [\%]& PWG+HER+UE [\%]  &PWG+PYT [\%] &PWG+PYT+MPI [\%]\\
\hline
\hline 
\\
$A^t_{\rm FB}$ total   &$-2.98\pm0.04$&$-2.95\pm0.05$&$-1.75\pm0.11$&$-1.70\pm0.11$&$-1.49\pm0.11$&$-1.36\pm0.11$\\ 
\\
$A^t_{\rm FB} \,,\ |y_t| < 1.0$   &$-2.60\pm0.04$&$-2.55\pm0.05$&$-1.51\pm0.12$&$-1.53\pm0.11 $&$-1.31\pm0.12$&$-1.22\pm0.12 $\\  
$A^t_{\rm FB} \,,\ |y_t| \ge 1.0$  &$-6.38\pm0.19$&$-6.51\pm0.15$&$-3.79\pm0.35$&$-3.15\pm0.34 $&$-2.99\pm0.35$&$-2.58\pm0.34 $\\
\\
$A^t_{\rm FB} \,,\ |m_{\ttb }| < 450~ \rm GeV$   &$-1.90\pm0.06$&$-1.80\pm0.06$&$-1.24\pm0.14$&$-1.24\pm0.13 $&$-0.81\pm0.14$&$-1.00\pm0.14 $\\
$A^t_{\rm FB} \,,\ |m_{\ttb }| \ge 450~ \rm GeV$  &$-4.70\pm0.06$&$-4.77\pm0.08$&$-2.70\pm0.19$&$-2.54\pm0.18 $&$-2.66\pm0.18$&$-1.98\pm0.18 $\\
\\
$A^{t}_{\rm FB}\,,\ \pt^{\ttb } \ge 10~ \rm GeV$ &$-2.95\pm0.04$&$-2.93\pm0.05$&$-2.64\pm0.06$&$-2.59\pm0.06 $&$-2.58\pm0.06$&$-2.39\pm0.06$\\
$A^{t}_{\rm FB}\,,\ \pt^{\ttb } \ge 20~ \rm GeV$ &$-2.41\pm0.05$&$-2.94\pm0.05$&$-2.80\pm0.05$&$-2.80\pm0.05 $&$-2.85\pm0.05$&$-2.81\pm0.05$\\
$A^{t}_{\rm FB}\,,\ \pt^{\ttb } \ge 35~ \rm GeV$ &$-3.90\pm0.06$&$-3.85\pm0.05$&$-3.54\pm0.06$&$-3.55\pm0.06 $&$-3.67\pm0.06$&$-3.63\pm0.06 $\\
$A^{t}_{\rm FB}\,,\ \pt^{\ttb } \ge 50~ \rm GeV$ &$-4.31\pm0.07$&$-4.33\pm0.06$&$-4.00\pm0.07$&$-4.02\pm0.07 $&$-4.19\pm0.07$&$-4.19\pm0.07 $\\\
$A^{t}_{\rm FB}\,,\ \pt^{\ttb } \ge 75~ \rm GeV$ &$-4.88\pm0.08$&$-4.62\pm0.08$&$-4.33\pm0.09$&$-4.29\pm0.09 $&$-4.59\pm0.09$&$-4.56\pm0.09 $\\
\\
\hline
\hline
\\
$A^\ttb_{\rm FB}$ total  &$-4.40\pm0.04$&$-4.34\pm0.05$&$-2.80\pm0.11$&$-2.54\pm0.11  $&$-2.22\pm0.11$&$-1.84\pm0.11 $\\ 
\\
$A^\ttb_{\rm FB} \,,\ |\Delta y_{\ttb}| < 1.0$  &$-2.70\pm0.04$&$-2.62\pm0.05$&$  -1.71\pm0.11$&$-1.91\pm0.11$&$-1.39\pm0.11$&$-1.16\pm0.11 $\\
$A^\ttb_{\rm FB} \,,\ |\Delta y_{\ttb}| \ge 1.0$ &$-19.48\pm0.18$&$-19.54\pm0.22$&$-10.52\pm0.52$&$-9.75\pm0.51 $&$-9.22\pm0.52$&$-7.54\pm0.51  $\\
\\
$A^\ttb_{\rm FB} \,,\ |m_{\ttb }| < 450~ \rm GeV$   &$-3.59\pm0.06$&$-3.51\pm0.06$&$-2.67\pm0.14$&$-2.36\pm0.13 $&$-1.74\pm0.14$&$-1.63\pm0.14 $\\
$A^\ttb_{\rm FB} \,,\ |m_{\ttb }| \ge 450~ \rm GeV$  &$-5.70\pm0.06$&$-5.66\pm0.08$&$-3.03\pm0.19$&$-2.88\pm0.18 $&$-3.06\pm0.18$&$-2.20\pm0.18 $\\
\\
$A^{\ttb}_{\rm FB}\,,\ \pt^{\ttb } \ge 10~ \rm GeV$  &$-4.35\pm0.04$&$-4.32\pm0.05$&$-3.98\pm0.06$&$-3.86\pm0.06  $&$-3.72\pm0.06$&$-3.51\pm0.06 $\\
$A^{\ttb}_{\rm FB}\,,\ \pt^{\ttb } \ge 20~ \rm GeV$  &$-3.71\pm0.05$&$-4.29\pm0.05$&$-4.22\pm0.05$&$-4.18\pm0.05 $&$-4.15\pm0.05$&$-4.11\pm0.05  $\\
$A^{\ttb}_{\rm FB}\,,\ \pt^{\ttb } \ge 35~ \rm GeV$  &$-5.72\pm0.06$&$-5.52\pm0.05$&$-5.16\pm0.06$&$-5.17\pm0.06 $&$-5.21\pm0.06$&$-5.21\pm0.06 $\\
$A^{\ttb}_{\rm FB}\,,\ \pt^{\ttb } \ge 50~ \rm GeV$  &$-6.25\pm0.07$&$-6.11\pm0.06$&$-5.70\pm0.07$&$-5.74\pm0.07 $&$-5.85\pm0.07$&$-5.92\pm0.07 $\\
$A^{\ttb}_{\rm FB}\,,\ \pt^{\ttb } \ge 75~ \rm GeV$  &$-6.62\pm0.08$&$-6.45\pm0.08$&$-5.99\pm0.09$&$-5.94\pm0.09 $&$-6.27\pm0.09$&$-6.25\pm0.09 $\\
\\
\hline
\hline
\end{tabular}

\caption{ \label{tab:yasym} Results for the forward-backward
  asymmetries $A^t_{\rm FB}$ and $A^\ttb_{\rm FB}$ at the Tevatron,
  for the $\ttb j$ sample with various acceptance cuts. Jets
  are reconstructed by the inclusive $\kt$-algorithm with $R=1$, 
above the $\pt^{\, \rm jet} > 20~\rm GeV$ minimum jet cut.}
\end{center}
\end{sidewaystable}

\subsubsection{Top-quark forward-backward  charge asymmetry}
\label{sec:afbtev}
In view of the measurements of the forward-backward asymmetry in
top-quark pair-production at the
Tevatron~\cite{Abazov:2007qb,Aaltonen:2008hc,Aaltonen:2011kc} we have
also evaluated the values of the $\ttb $ rapidity asymmetries in $t \tb
+$jet samples, at various stages of our simulations and with various
acceptance cuts.  The definitions for $A^t_{\rm FB}$ and $A^{\ttb}_{\rm FB}$ used throughout this paper 
are 
\beq
\label{eq:afbt}
A^t_{\rm FB} = \frac{1}{\sigma} \left(\ {\displaystyle \int\limits_{\scriptscriptstyle
      y_t>0}} d\sigma - {\displaystyle \int\limits_{\scriptscriptstyle
      y_t<0}} d\sigma \right) \,, \qquad
A^{\ttb}_{\rm FB} = \frac{1}{\sigma} \left(\ {\displaystyle
    \int\limits_{\scriptscriptstyle \Delta y_{\ttb}>0}} d\sigma -
  {\displaystyle \int\limits_{\scriptscriptstyle \Delta y_{\ttb}<0}}
  d\sigma \right)\,, \eeq
where $\Delta y_{\ttb}= y_t-y_{\tb}$.  Due to the fact that the
initial $p\bar p$ state at the Tevatron is a CP eigenstate and due to
the absence of CP violating effects in QCD,
one has $\frac{d\sigma}{d y_{t}} (y_t) = \frac{d\sigma}{d y_{\tb}}
(-y_{\tb})$ and $A^{\ttb}_{\rm FB}$ corresponds to $A^t_{\rm FB}$ in
the $\ttb$ rest frame. When moving to the laboratory frame, where a
greater accuracy can be reached on the experimental determination of
these quantities, $A^t_{\rm FB}$ gets smaller in magnitude than
$A^{\ttb}_{\rm FB}$, which in turn is a boost-invariant quantity,
depending only on rapidity differences.  We report both in
Tab.~\ref{tab:yasym}, expressed as percentage values to ease their
readability, at various stages of our simulation. We have also
included the predictions obtained with the \HERWIG{} and \PYTHIA{}
showers supplemented with underlying-event and multi-parton
interactions activities, respectively, referring to them as {\tt
  PWG+HER+UE} and {\tt PWG+PYT+MPI}. In case of {\tt PWG+PYT+MPI} the
Perugia~0 Tune~\cite{Skands:2010ak} -- {\tt MSTP(5)=320} -- has also
been adopted.

For both asymmetries and shower models, we observe that the inclusion
of the parton shower leads to a significant correction compared to the
fixed order and LHEF predictions.  For example, the results for total
asymmetries are almost reduced by a factor of two by the
parton-shower.  A similar effect is also observed in presence of
selection cuts on the rapidity or invariant mass of the heavy quark
pair. This outcome extends the conclusions obtained by previous
combined NLO and parton shower simulations~\cite{Aaltonen:2011kc} for
the inclusive $\ttb $ sample.\footnote{ See also~\cite{Ahrens:2011uf}
  for higher-order QCD corrections to $A_{FB}$ and
  ~\cite{Bevilacqua:2011hy} for studies of the $\ttb +2$jet process in
  particular.}

 Inspecting the results shown in Tab.~\ref{tab:yasym} more closely, we
 find that performing \PYTHIA{} and \HERWIG{} showers leads to similar
 predictions for $A^t_{\rm FB}$: the values are fairly consistent once
 the statistical uncertainties are taken into account. The agreement
 is instead much worse for $A^{\ttb}_{\rm FB}$.  The inclusion of
 underlying-event or multi-particle interaction does not dramatically
 change this picture, the interleaved shower evolution due to MPI and
 Perugia~0 Tune in \PYTHIA{} results  in a larger difference with
 respect to the default behaviour than the simpler UE model in
 \HERWIG{}.

The dependence of $A_{\rm FB}$ on the transverse momentum of the
$t\tb$-pair, $\pt^{\ttb }$, has also recently been reported
in~\cite{Abazov:2011rq}. In ref.~\cite{Kuhn:2011ri}, the authors have
investigated this behaviour in presence of selection cuts on the
$\ttb$ pair invariant mass, including also EW corrections. 
\clearpage Since this
observable only receives contributions starting at ${\cal O} \(
\as^3\)$, even for the $\ttb $ inclusive sample -- an extra jet is
always required to have a non-zero $\pt^{\ttb }$ -- our approach offers
the possibility to evaluate it at NLO supplemented with shower effects
for the first time.  This should improve over the \MCatNLO{} and
\POWHEG{} simulations for $\ttb$ production, that are only
leading-order/leading-logarithm accurate for $A_{\rm FB}$ when
$\pt^{\ttb} > 0$.

In Tab.~\ref{tab:yasym} we have also included our predictions for the
forward-backward asymmetries with $\pt^{\ttb }>0$. We now see, that
the results after showering with two different SMCs are consistent
almost everywhere, apart from the very low $\pt^{\ttb}$ region, which
thus contributes to generate the differences in $A^\ttb_{\rm FB}$.
Moreover, for both asymmetries $A^t_{\rm FB}$ and $A^\ttb_{\rm FB}$,
we observe that for $\pt^{\ttb } > 10~\rm GeV$ the inclusion of the
parton-shower leads to results in fair agreement with the NLO ones.
The large differences between NLO and showered results observed in the
total asymmetries are thus traced back to the region of small
$\pt^{\ttb }$'s. This appears plausible since in that region soft and
collinear emission produced by the parton shower will be more
important compared to the large $\pt^{\ttb }$ region. We also note
that our findings are different from the observations made
in~\cite{Abazov:2011rq} where a stronger effect of the parton shower
using specific tunes was observed over the entire $\pt^{\ttb }$ range.
One has to remind, however, that for a leading-order SMC such as the
\PYTHIA{} program used in~\cite{Abazov:2011rq}, the contributions to
the asymmetry for $\pt^{\ttb} > 0$ are generated exclusively by the
shower. It thus looks feasible that a change in the parton shower will
be reflected in a overall change of the predictions.

Whether the aforementioned large corrections produced by the parton
shower could be responsible in part for the observed discrepancy
between theory and experiment is however unclear because in the $\ttb$
sample this effect may cancel. Given the stability of the theoretical
prediction above $ \pt^{\ttb }=10$ GeV ---now available at NLO
accuracy--- a detailed comparison of experimental data and theory
predictions may provide useful information to further scrutinize the
observed discrepancy in the inclusive sample.


\subsection{LHC}
\label{sec:res-lhc}

\begin{figure}[tb]
\centering
\vspace*{10mm}
    {
    \includegraphics[width=\figwidth]{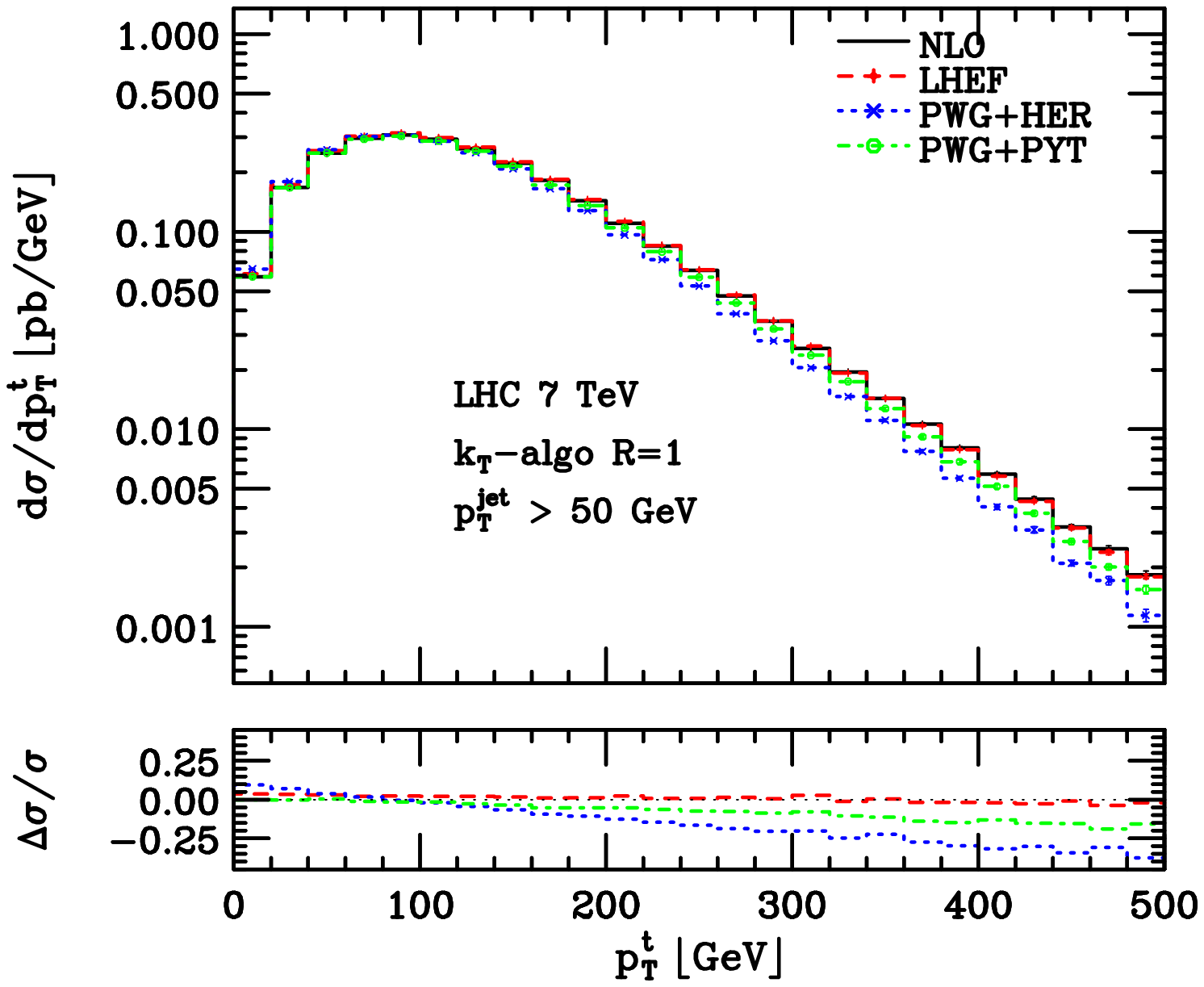}
    \hfill
    \includegraphics[width=\figwidth]{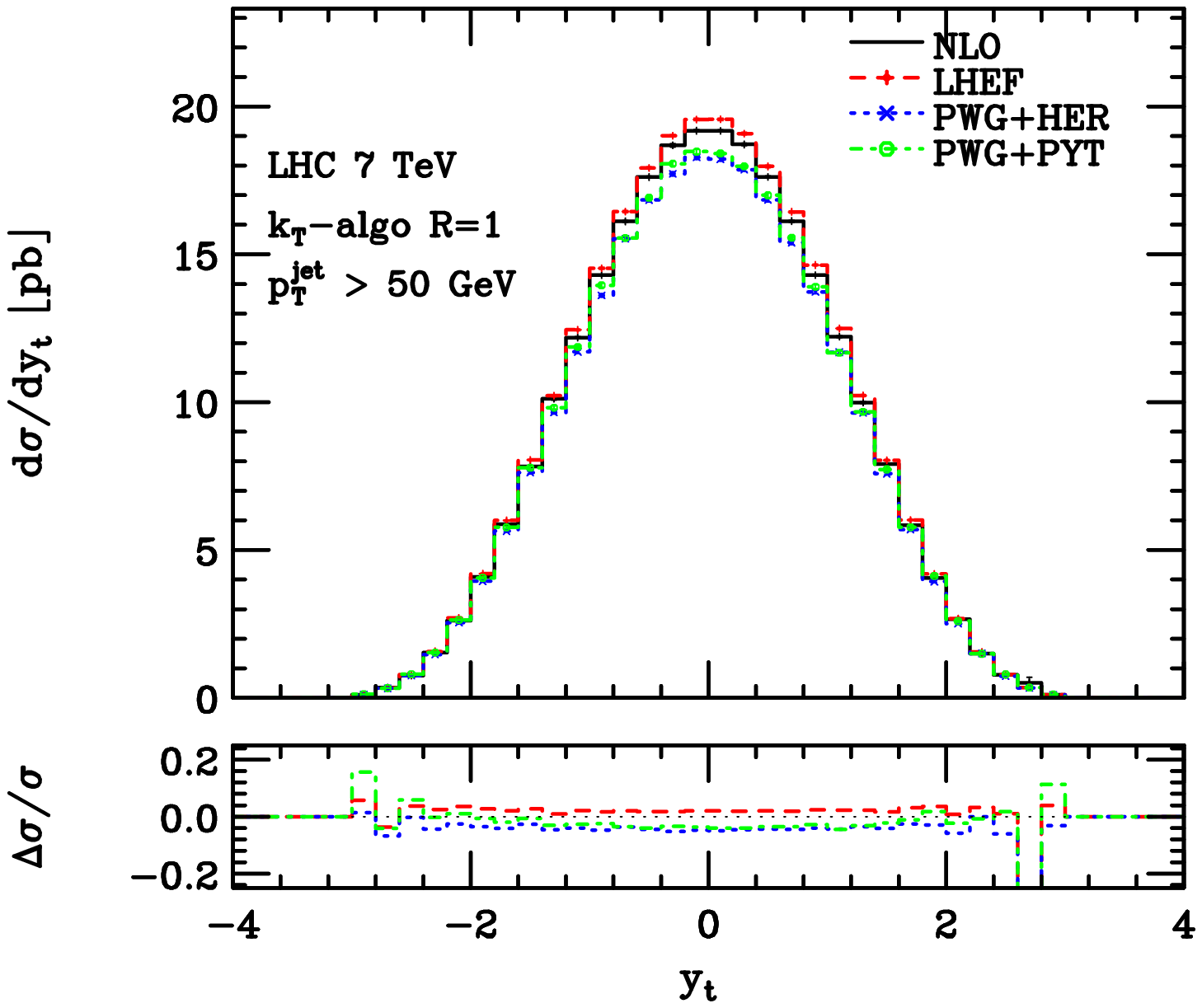}
    }
    \caption{ \small
      \label{fig:LHC-pt_t-y_t}
The differential cross sections as function of the transverse momentum
$\pt^{\;t}$ and of the rapidity $y_t$ of the top-quark at the LHC ($\sqrt{s}=7$~TeV).  }
\end{figure}

\begin{figure}[tb]
\centering
\vspace*{10mm}
    {
    \includegraphics[width=\figwidth]{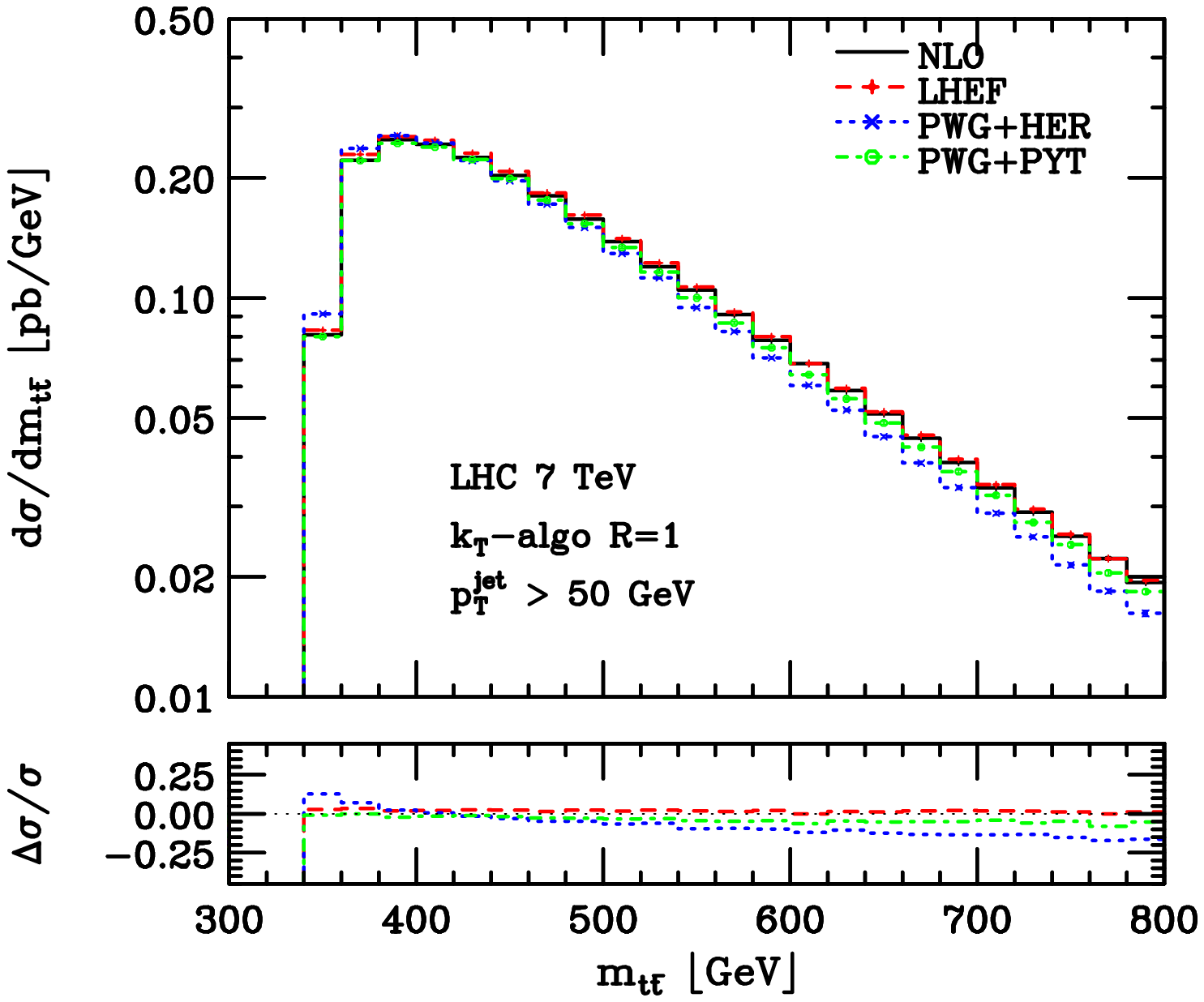}
    \hfill
    \includegraphics[width=\figwidth]{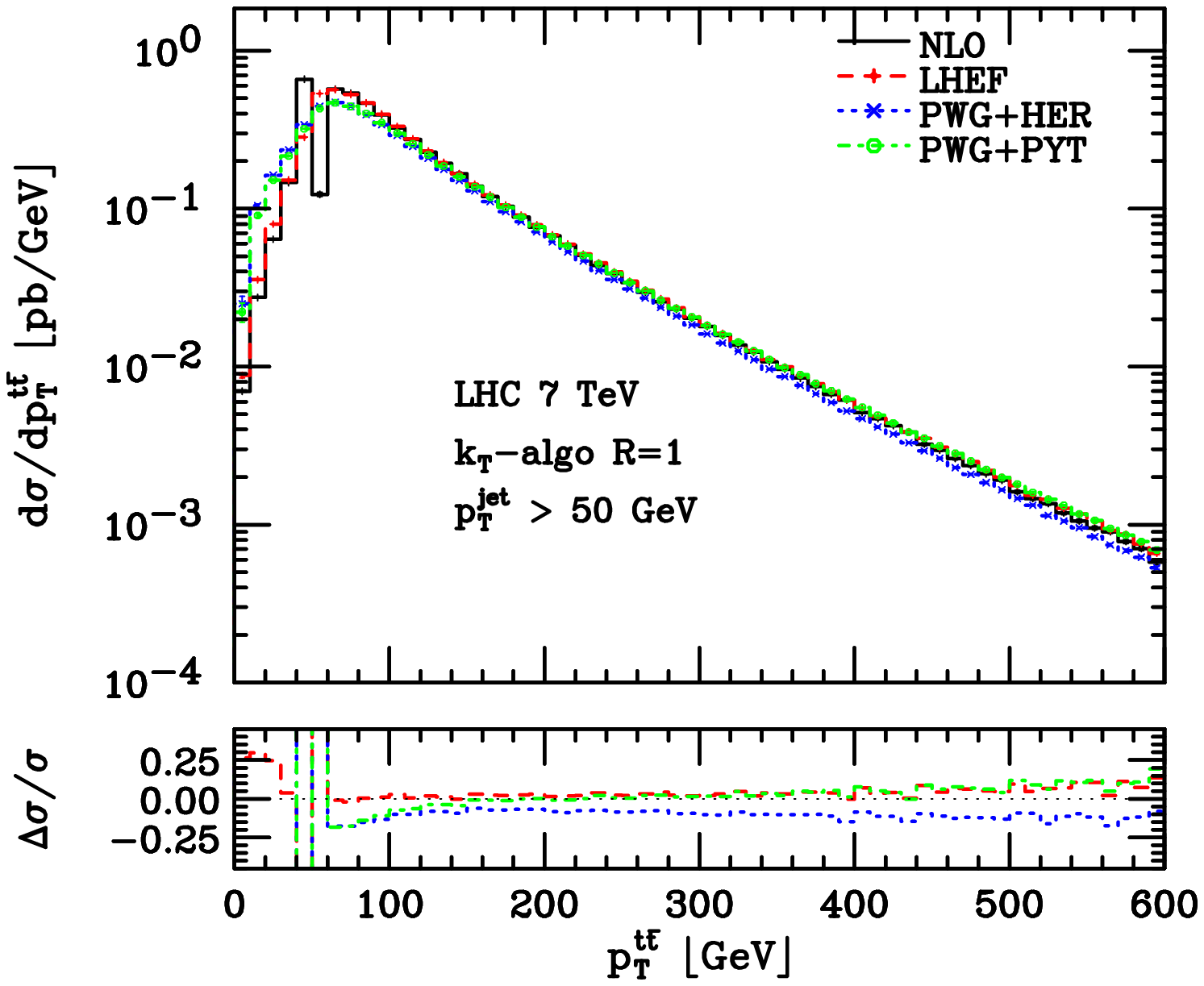}
    }
    \caption{ \small
      \label{fig:LHC-mtt-pttt}
The differential cross sections as function of the $\ttb$-pair
invariant mass $m_{\ttb}$ and transverse momentum $\pt^{\ttb}$ at the
LHC ($\sqrt{s}=7$~TeV).  }
\end{figure}

\begin{figure}[tb]
\centering
\vspace*{10mm}
    {
    \includegraphics[width=\figwidth]{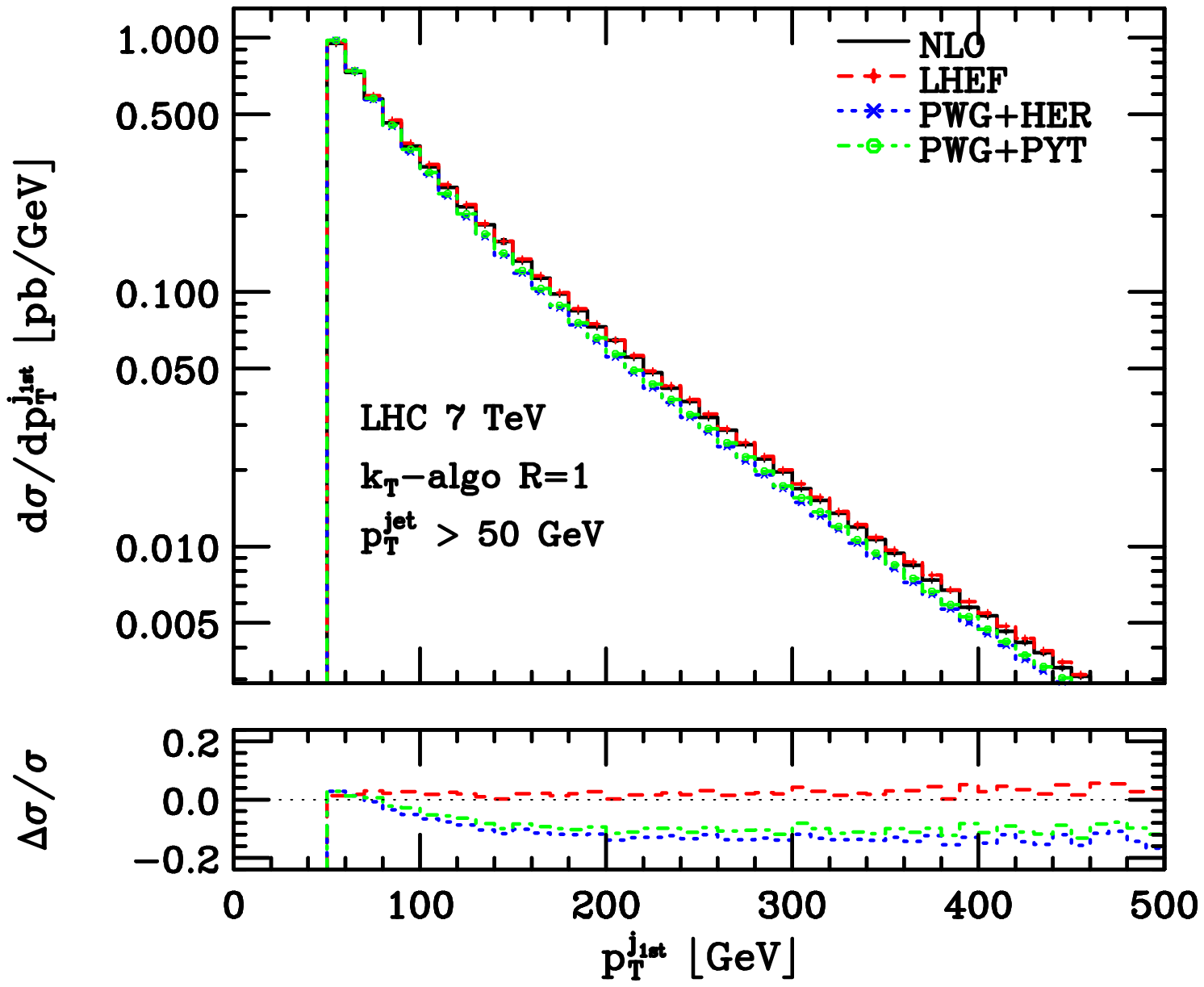}
    \hfill
    \includegraphics[width=\figwidth]{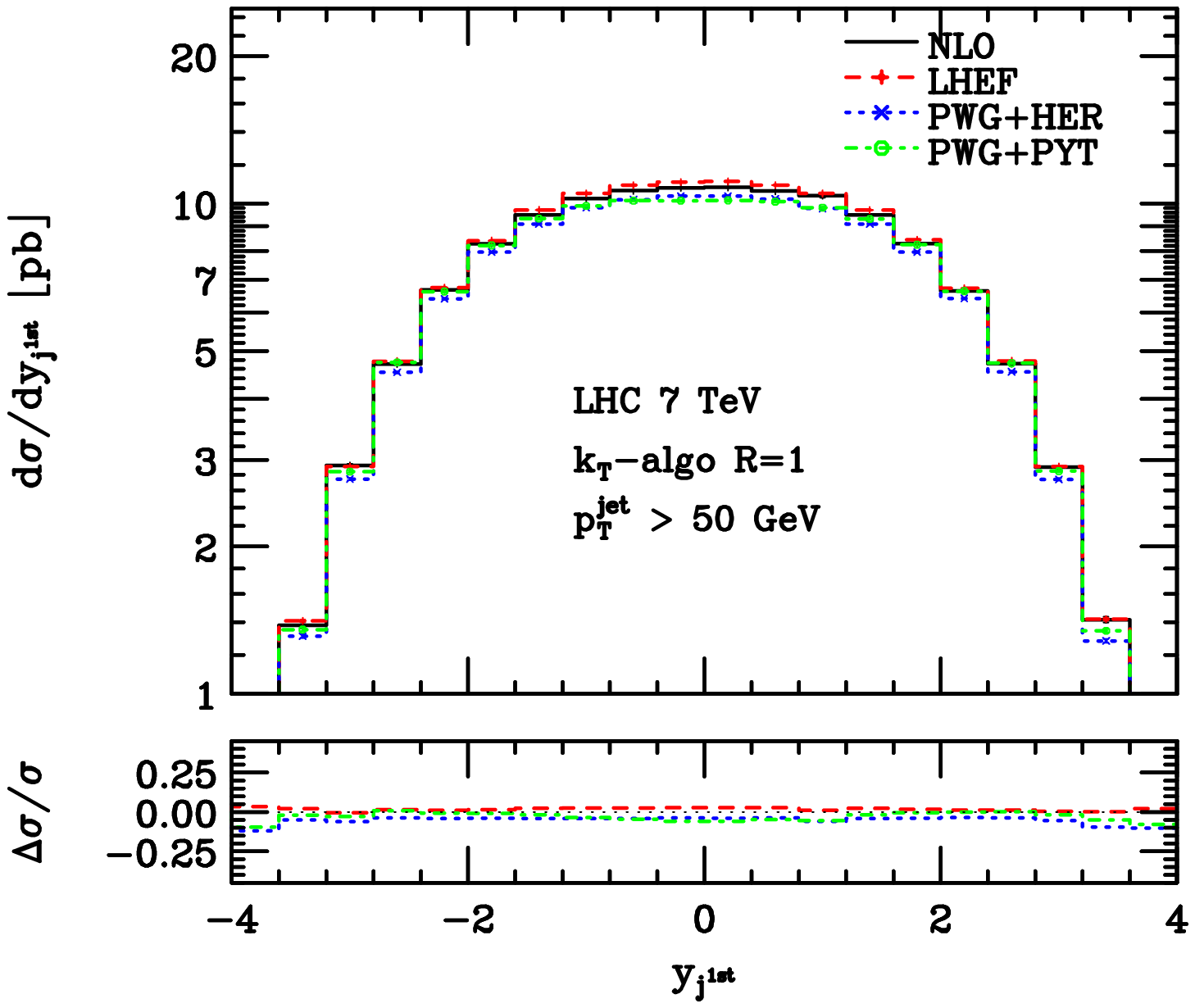}
    }
    \caption{ \small
      \label{fig:LHC-j1_pt-j1_y}
The differential cross sections as function of the hardest jet
transverse momentum and rapidity at the LHC ($\sqrt{s}=7$~TeV).  }
\end{figure}

\begin{figure}[tb]
\centering
\vspace*{10mm}
    {
    \includegraphics[width=\figwidth]{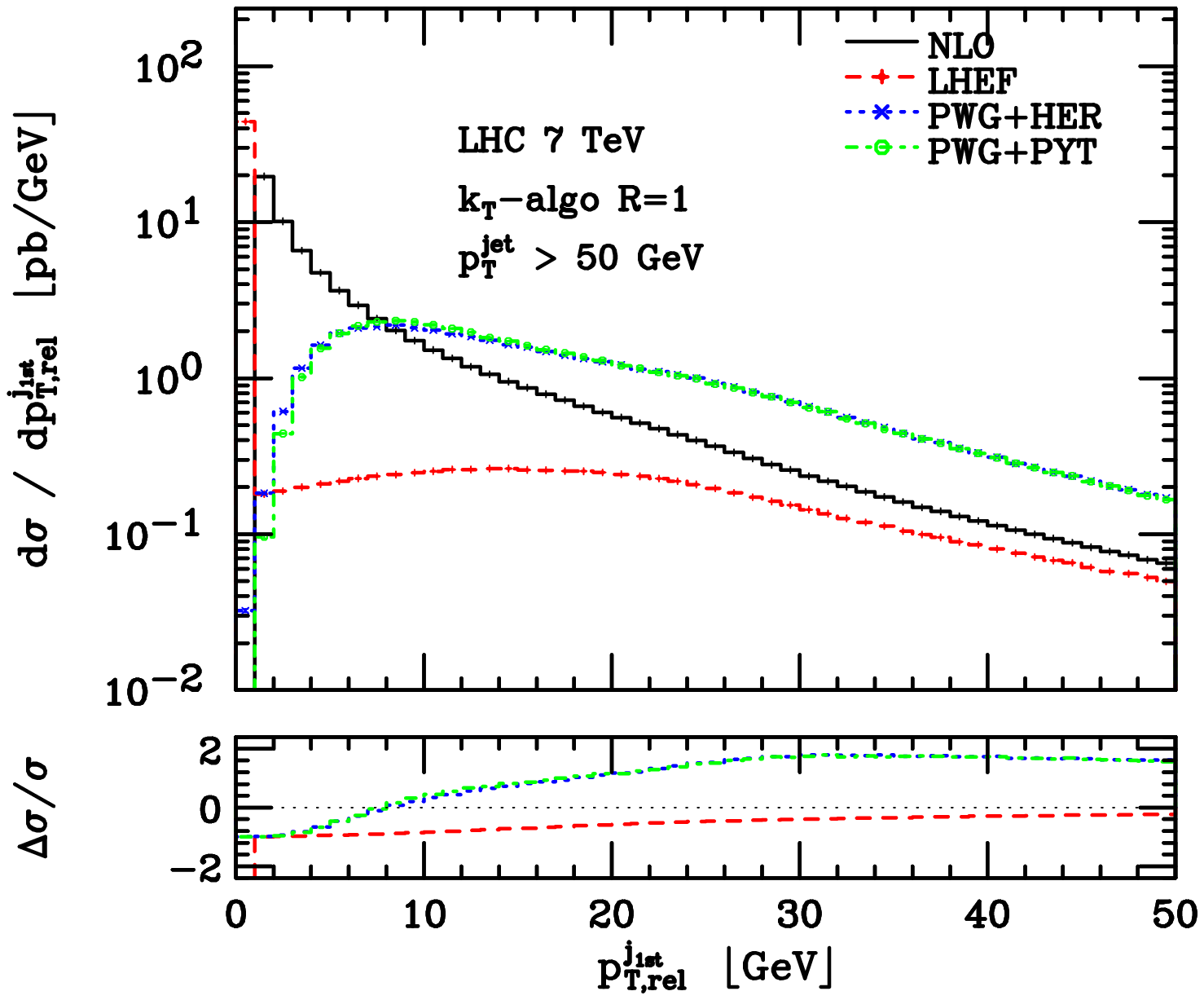}
    \hfill
    \includegraphics[width=\figwidth]{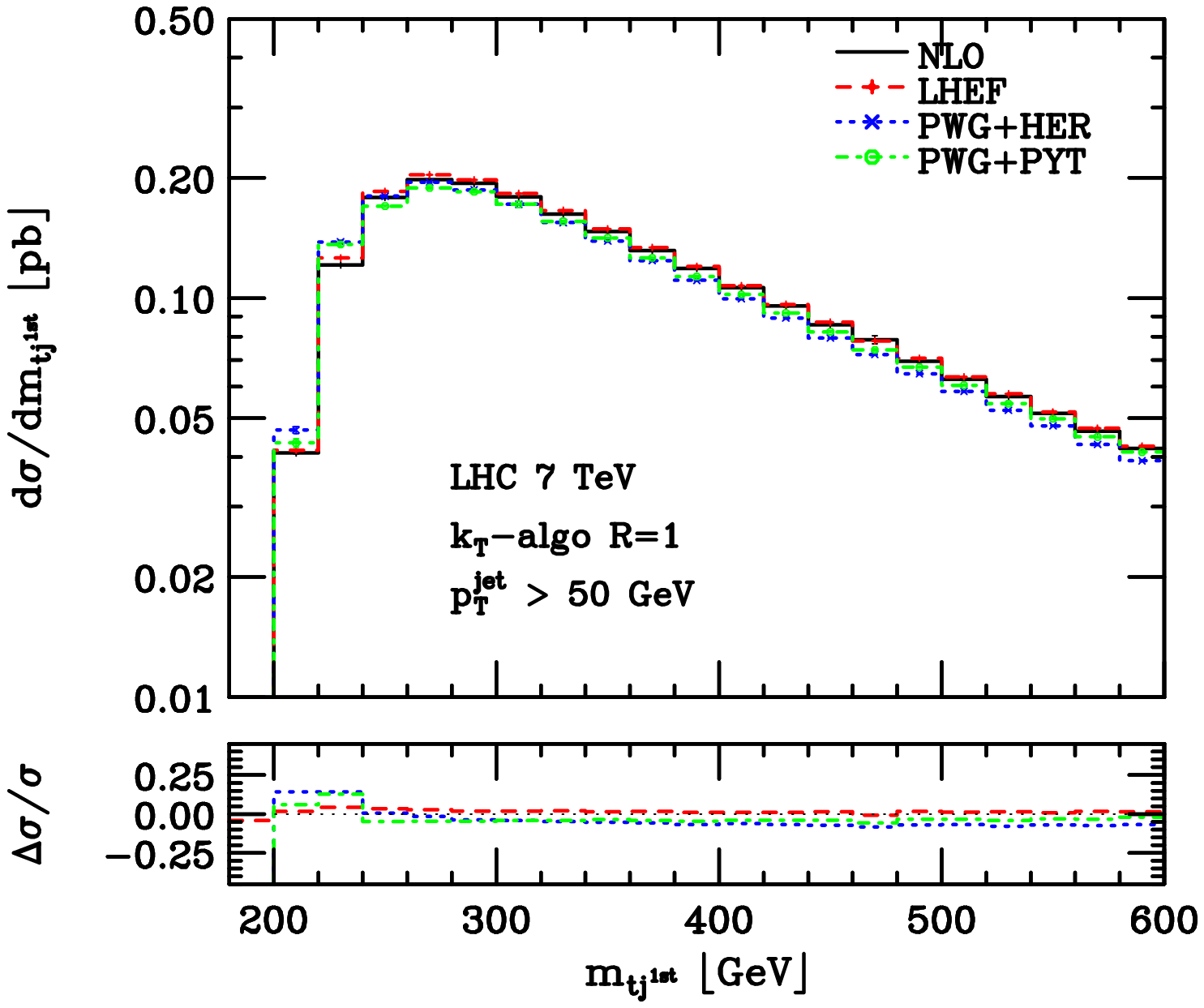}
    }
    \caption{ \small
      \label{fig:LHC-j1_ptrel-tj1_invm}
The differential cross sections as function of the scalar sum of
transverse momenta inside the hardest jet (\figleft{} panel) , as
defined in eq.~(\ref{eq:ptrel}), and the invariant mass of the system
made by the top-quark and the hardest jet (\figright{} panel), at the
LHC ($\sqrt{s}=7$~TeV).  }
\end{figure}

\begin{figure}[tb]
\centering
\vspace*{10mm}
    {
    \includegraphics[width=\figwidth]{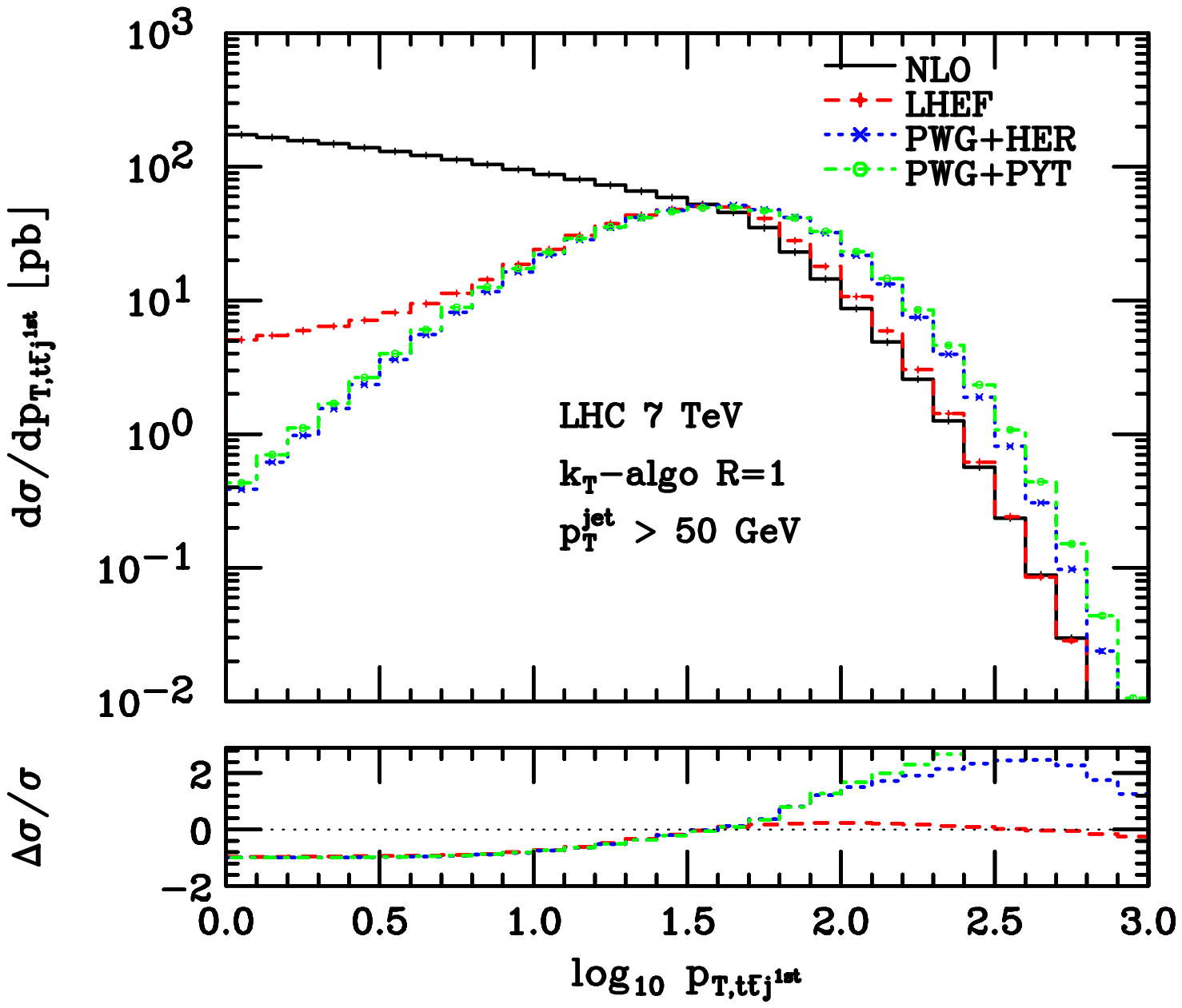}
    \hfill
    \includegraphics[width=\figwidth]{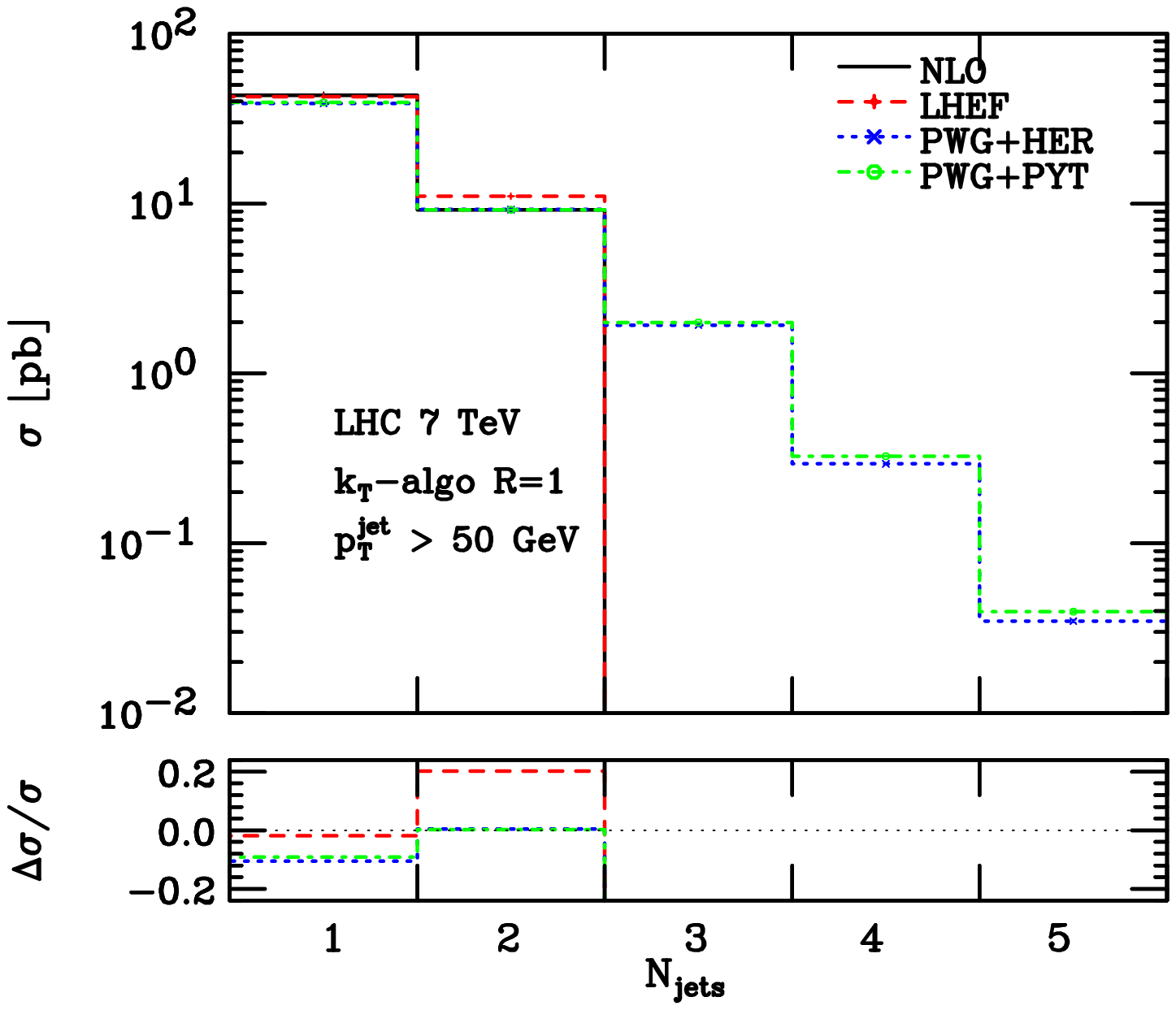}
    }
    \caption{ \small
      \label{fig:LHC-ttbarj1_logpt-njets}
The differential cross sections as function of the transverse momentum
of the system made by $\ttb$-pair and the hardest jet (\figleft{} panel) and 
the total inclusive cross
section for $\sigma_{\ge N_{\rm jets}}$ (\figright{} panel) at the LHC 
($\sqrt{s}=7$~TeV).}
\end{figure}

\begin{figure}[tb]
\centering
\vspace*{10mm}
    {
    \includegraphics[width=\figwidth]{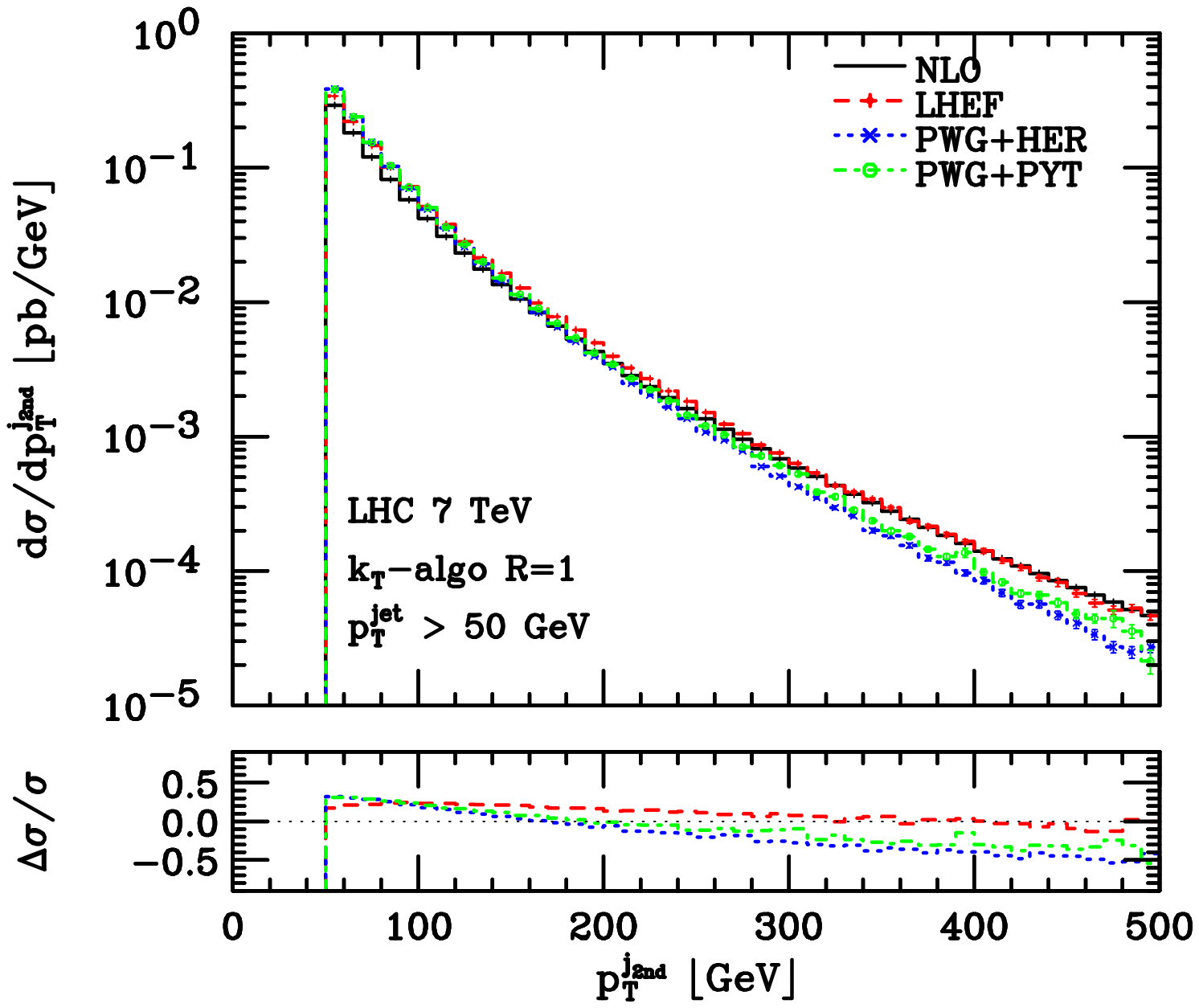}
    \hfill
    \includegraphics[width=\figwidth]{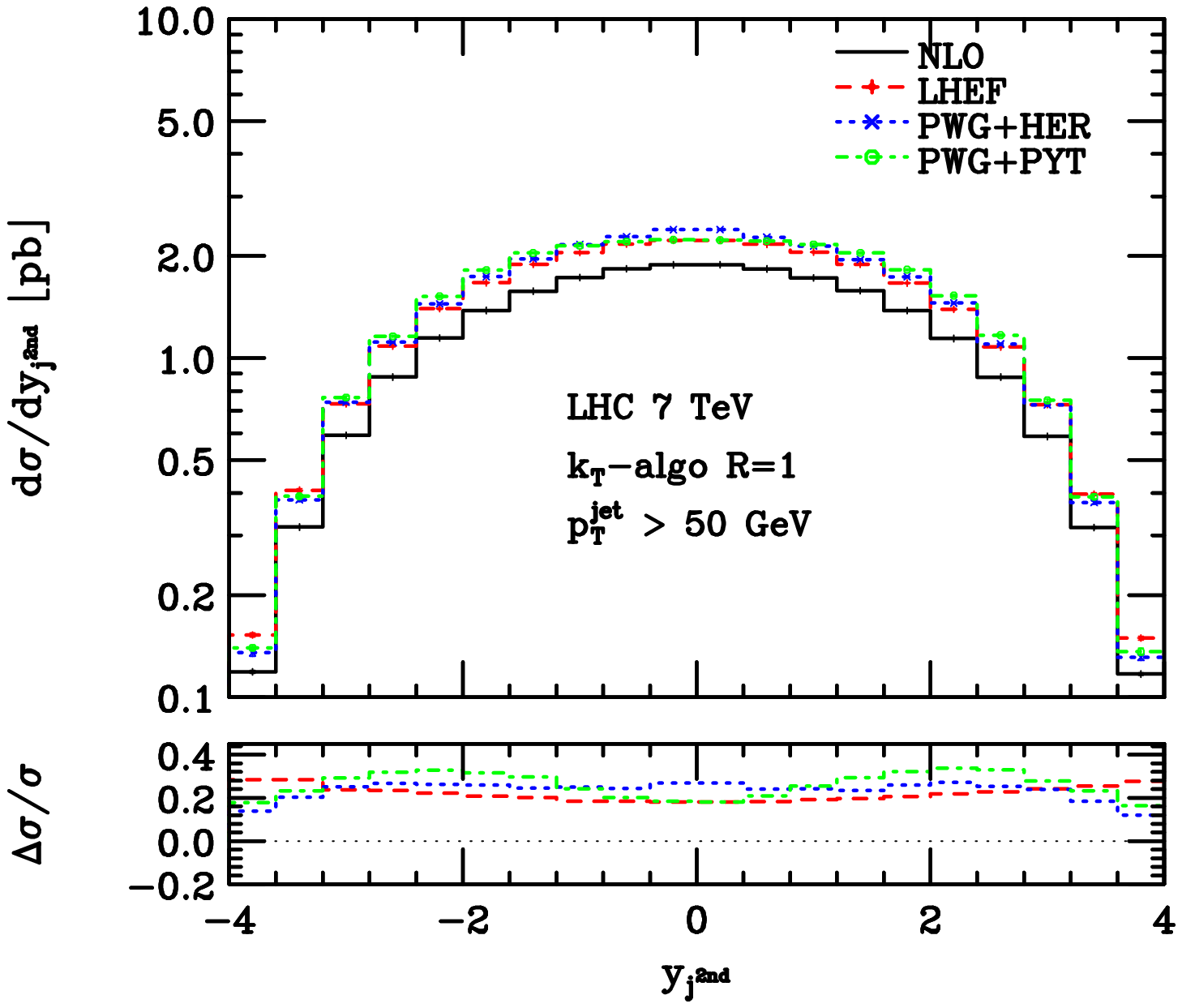}
    }
    \caption{ \small
      \label{fig:LHC-j2_pt-j2_y}
The differential cross sections as function of the next-to-hardest jet
transverse momentum and rapidity at the LHC ($\sqrt{s}=7$~TeV).  }
\end{figure}

In Figs.~\ref{fig:LHC-pt_t-y_t} to~\ref{fig:LHC-j2_pt-j2_y} we report
a similar set of distributions as already shown for the Tevatron, this
time for the LHC collider at $\sqrt{s} = 7$~TeV.  For most of the
observables, the same comments and conclusions as reported for the
Tevatron case are, in general, valid here, too. For
this reason, we will not repeat the same
remarks here. There are, however, some new plots that deserve
discussion and explanations.
%

In the \figright{} panel of Fig.~\ref{fig:LHC-pt_t-y_t} we show the
differential cross section as a function of the rapidity
distribution of the top-quark $y_t$. This inclusive quantity shows
the expected good agreement between NLO and LHEF results.  It is
also pretty stable with respect to the inclusion of parton shower
effects, resulting  only in a few percent change in
the overall normalization.

In the \figleft{} panel of Fig.~\ref{fig:LHC-j1_ptrel-tj1_invm} we show
the differential cross section $d\sigma/d \ptrel^{j_{1\rm st}}$ as
function of the scalar sum of the relative transverse momenta of the
particles in the first jet, $\ptrel^{j_{1\rm st}}$, which is defined with
respect to the jet axis in the frame where the jet has zero rapidity.
In general, we define for the $n$-th jet,
\begin{equation}
\label{eq:ptrel}
\ptrel^{j_{n\rm th}} = 
\sum_{i \in j_{n\rm th}} \frac{| \vec{k}_i \times \vec{p}^{j_{n\rm th}}|}{|\vec{p}^{j_{n\rm th}}|}
\, ,
\end{equation}
where $k_i$ denotes the momentum of the $i^{\rm th}$ particle in the
$n$-th jet, see also~\cite{Alioli:2010xa}.  
The observable behaves as expected: The NLO curve shows the unphysical
enhancement associated with the IR singularity at $\ptrel^{j _{1\rm st}}=0$,
the LHEF result corrects for this only partially, owing to the Sudakov suppression
by final-state radiation, while the fully showered events manifest the
physical suppression at low $\ptrel^{j_{1\rm st}}$.
Moreover, in the \figright{} panel of
Fig.~\ref{fig:LHC-j1_ptrel-tj1_invm} we show the distribution of the invariant
mass of the system made by the top-quark and the hardest light jet.

Eventually, in the \figright{} panel of
Fig.~\ref{fig:LHC-ttbarj1_logpt-njets}, the total exclusive cross
section as a function of the number of resolved jets, $N_{jets}$, is
shown.  The moderate difference in the $N_{jets}=2$ bin between the
NLO and LHEF results may be considered a consequence of the different
normalization of the distributions involving the next-to-hardest jet,
as shown {\it e.g.}, in Fig.~\ref{fig:LHC-j2_pt-j2_y}.  Higher bins
are instead populated only by the shower and thus predictions for
those bins are only accurate in the strict soft/collinear limit.

One noticeable difference w.r.t. Tevatron results is the reduced
discrepancy between the showered and partonic results for observables
that involve the hardest or next-to-hardest jets, {\it e.g.},
Figs.~\ref{fig:LHC-j1_pt-j1_y} and~\ref{fig:LHC-j2_pt-j2_y}.  A
possible explanation for this is in the higher jet cut ($\pt =
50$~GeV) used for LHC analysis, that largely restricts the regions
where shower effects may become important.

A first phenomenological application of the \POWHEG{} implementation discussed
here has already been reported in~\cite{lpposter:2011xx}, namely 
the study of $\ttb j$-events at the LHC as a function of the invariant mass of the
multi-jet systems as a means of measuring the top-quark mass.

\begin{sidewaystable}[tb]
\begin{center}
\begin{tabular}{ccccccc}
\hline
\hline
LHC $7$~TeV &NLO [\%]&LHEF [\%]&PWG+HER [\%]&PWG+HER+UE [\%]&PWG+PYT[\%]&PWG+PYT+MPI [\%]\\
\hline
\hline
\\ 
$A^\eta_{C} $ &$0.19\pm 0.09 $ &$ 0.18 \pm 0.06 $&$ 0.46 \pm 0.10  $&$ 0.26 \pm 0.11 $&$ 0.40 \pm 0.11 $&$0.57 \pm 0.11 $\\ 
$A^y_{C} $   &$0.51\pm 0.09 $  &$ 0.47\pm 0.06 $&$ 0.73 \pm 0.10 $&$ 0.52 \pm 0.11 $&$ 0.66 \pm 0.11 $&$ 0.76 +\- 0.11 $\\ 
\\
\hline
\hline
\end{tabular}
\end{center}
\caption{
\label{tab:chasym}
 Results for the top-quark charge asymmetry $A_C$ in the $\ttb  j$ sample at
 the LHC ($\sqrt{s} = 7$~TeV). Jets
  are reconstructed by the inclusive $\kt$-algorithm with $R=1$, 
above the $\pt^{\, \rm jet} > 50~\rm GeV$ minimum jet cut.}

\vspace{2cm}
\begin{center}
\begin{tabular}{cccccc}
\hline
\hline
 Tevatron $1.96$~TeV &LHEF [\%]&PWG+HER [\%]&PWG+HER+UE [\%]&PWG+PYT [\%]&PWG+PYT+MPI [\%]\\
\hline
\hline 
\\
$A^{\ell^+}_{\rm FB} $ &$-1.83 \pm 0.05$&$ -0.94 \pm 0.11 $&$ -0.92 \pm 0.11$&$ -0.57 \pm 0.11 $  & $-0.41 \pm 0.11 $\\ 
$A^{\ell^+ \ell^-}_{\rm FB} $ &$ -2.21 \pm 0.05 $&$-1.00 \pm 0.11 $&$-0.99 \pm 0.11 $&$-0.59 \pm 0.11 $&$-0.46 \pm 0.11 $\\ 
\\
\hline
\hline
& & PWG+HER' [\%]&PWG+HER'+UE [\%]&PWG+PYT' [\%]&PWG+PYT'+MPI[\%]\\
\hline
\hline 
\\
$A^{\ell^+}_{\rm FB} $ &$ $&$ -0.69 \pm 0.11 $&$ -0.93 \pm 0.11$&$ -0.71 \pm 0.11 $  & $-0.47 \pm 0.11 $\\ 
$A^{\ell^+ \ell^-}_{\rm FB} $ &$  $&$-1.13 \pm 0.11 $&$-1.11 \pm 0.11 $&$-0.90 \pm 0.11 $&$-0.85 \pm 0.11 $\\ 
\\
\hline
\hline
\end{tabular}
\end{center}
\caption{
\label{tab:yasym3}
 Results for the lepton charge asymmetries $A^{\ell^+}_{\rm FB}$  and $A^{\ell^+ \ell^-}_{\rm FB}$ in the $\ttb  j$ sample at
 the Tevatron ($\sqrt{s}~=~1.96$~TeV).  Jets
  are reconstructed by the inclusive $\kt$-algorithm with $R=1$, 
above the $\pt^{\, \rm jet} > 20~\rm GeV$ minimum jet cut. }

\end{sidewaystable}

\subsubsection{Top-quark charge asymmetry}
\label{sec:aclhc}
In Tab.~\ref{tab:chasym} we present our estimate of the charge
asymmetries in the $\ttb j$ sample, measured with respect to the
(anti-) top-quark pseudo-rapidity $\eta$ and rapidity $y$, at various
stages of our simulation.  The two observables are defined in the same
way \beqn
\label{eq:ach}
A^{x}_C &=& \frac{1}{\sigma} \left( \ {\displaystyle
    \int\limits_{\scriptscriptstyle \Delta{|x|}>0}} d\sigma -
  {\displaystyle \int\limits_{\scriptscriptstyle \Delta{|x|}<0}}
  d\sigma \right) \,, \eeqn

with $\Delta{|x|} = |x_t| - |x_{\tb}|$ and $x=\eta$ or $x=y$,
respectively.  By looking at our results, one can see that a positive
asymmetry is present for both observables, albeit larger in magnitude
for $A^{y}_C$.  Moreover, results obtained with the inclusion of
\PYTHIA{} and \HERWIG{} showers are in fair agreement, even after the
addition of MPI or UE.  We also observe that the numerical evaluation
of the asymmetry is rather challenging at fixed order, leading to
large statistical uncertainties. Only after increasing the statistics
by one order of magnitude we were able to obtain errors comparable
with those obtained after the inclusion of the parton shower. We
consider the results including the effects of the parton shower as the
best prediction currently available for the asymmetry.  We refer to
refs.~\cite{ATLAS-CONF-2011-106,CMS-PAS-TOP-11-014} for recent LHC
measurements of $A^{y}_C$ and $A^{\eta}_C$ in the $\ttb $ inclusive
sample.
\\

In closing we remark, that we have also carefully tested the LHC
nominal center-of-mass energy, producing the
same set of distributions as in Figs.~\ref{fig:LHC-pt_t-y_t}
to~\ref{fig:LHC-j2_pt-j2_y} for $\sqrt{s} = 14$~TeV collisions.  Since no
significant difference in the behaviour  has been
observed in any of the distributions, with respect to the $7$~TeV case, we refrain from including the
corresponding plots here.
\clearpage

\subsection{Top-quark decay}
\label{sec:topdecay}

The studies so far have been based on the assumption of stable
(anti-)top-quarks during both the fixed order calculation stage and
the generation of the hardest radiation according to the \POWHEG{}
method.  Thus, the output resulting from these stages includes
on-shell top-quarks and extra light partons only.  When interfacing to
a SMC program for performing the rest of the showering and the
hadronization steps, no information concerning the spin or helicity of
these top-quarks is retained and the decay is performed by the SMC,
which averages over the possible top polarizations.

In this way one does not correctly account for top-quark off-shell
effects, non-resonant production mechanisms and for the top-quark spin
correlations.  However, given the large ratio between the top-quark
mass and the top-quark width, the first two effects are suppressed as
${\cal O}(\Gamma_t/m_t)$ and, therefore, are rather modest. This was
also explicitly shown for the $\ttb $ production case by the recent
calculation of the NLO corrections to the $W^+W^-b\bar{b}$ final
state~\cite{Denner:2010jp, Bevilacqua:2010qb}.

Top-quark spin correlations at hadron colliders were instead realized
to be an important tool to study top-quark properties since 
the early work of~\cite{Bernreuther:2001rq}. 
Also, the QCD corrected spin analyzing power of jets in decays of polarized top-quarks has been 
extensively discussed in~\cite{Brandenburg:2002xr}. 
The possibility to measure these correlation effects is particular to top-quarks
because of their large mass and short lifetime, 
which prevent non-perturbative QCD effects 
from significantly depolarizing them before the decay.

The usual approach to study this phenomenon relies on taking the top-quark
zero-width limit and separately considering NLO corrections to the
production and decay processes. This is done by means of a spin
correlation matrix which accounts for all the possible spin
configurations occurring in the two stages. 
Several studies following this approach have already been presented, concerning both $\ttb $
production~\cite{Bernreuther:2001rq} with spin correlations at NLO and
the $\ttb  j$ case~\cite{Melnikov:2010iu} with spin correlations accounted for at leading order.

Here we follow a different approach, that was first introduced
in~\cite{Frixione:2007zp} and has already been employed and discussed
in both implementations, \POWHEG{}~\cite{Alioli:2009je,Re:2010bp,Oleari:2011ey} and
\MCatNLO{}~\cite{Frixione:2005vw,Frixione:2008yi}.  In
this way we avoid having to introduce the spin correlation matrices, but we
nevertheless recover the exact leading-order spin correlations.  To be
more precise, we have LO spin-correlations in the soft and collinear
regions, while the same accuracy of the complete NLO calculation,
namely the matrix element, 
\begin{equation}
  \label{eq:ME-ttbar-decay}
  pp (p\bar{p}) \to (t \to b \ell^+ \nu) (\tb  \to \bar{b} \ell^-
  \bar{\nu}) j j
  \, ,
\end{equation}
is used for the hard regions.\footnote{In case no radiation harder
  than $\ptmin=0.8$~GeV is generated by the \POWHEG{} method, the
  event is classified as Born-like: the corresponding matrix elements
  for the decayed process , {\it i.e.}, $pp (p\bar{p}) \to (t \to b
  \ell^+ \nu) (\tb  \to \bar{b} \ell^- \bar{\nu}) j $,
  are then employed.}  
For the sake of brevity, we explain this method here
only briefly, and instead, refer to the aforementioned publications
for more details.  The basic idea is to first
generate events with stable top-quarks (un-decayed events) through the
usual \POWHEG{} machinery and then generate the decay products
according to the matrix element for the full production and decay
process (decayed events). 

Moreover, in order to produce more realistic final states, we have
also allowed for a reshuffling of the momenta of the top-quark decay
products, resulting in off-shell top-quarks and $W$ bosons, whose
virtualities have been distributed according to Breit-Wigner shapes.
This reshuffling is such that the relative 3-momentum of the
(anti-)top-quark in the $\ttb $ rest frame and the
3-momentum of the radiated light partons in the partonic
center-of-mass system are kept fixed. It also takes
 care of the small changes in the phase space and
luminosity due to the $t$ and $\tb $ quark being slightly off-shell
with respect to the chosen pole mass value. In any case, we restrict
the top-quark and $W$ virtualities off-shellness to be
in between $10$ widths. We have checked that this modification does
not significantly change the results obtained in the top-quark and
$W$ boson zero-width approximation, by comparing several
distributions for the two cases.\footnote{ The zero-width
approximation in the decay may always
  be enforced by setting the {\tt zerowidth} token to 1 in the input
  card.}

Furthermore, since charged leptons (or down-type quarks) coming from the 
top-quark decays are the most sensitive probe of the
top-quark spin directions, 
we concentrate on the di-leptons channel only. 
We thus always assume 
$t\to W^+ b \to \ell^+ \nu b$ 
and 
$\tb \to W^- \bar{b} \to \ell^- \bar{\nu}_\ell \bar{b}$, 
with branching ratio $BR(W\to\ell \nu)=0.108$.  
However, we stress that our implementation is completely general for what concerns the top-quark decay products,
such that the lepton+jets or the all hadronic search channels can be
simulated as well, if desired.\footnote{See {\it e.g.},
  \cite{Mahlon:2010gw} for a review of observables sensitive to
  spin-correlation effects in the lepton+jets channel.}

Thus, taking as an example the semi-leptonic
top-quark decay channel, the conditional probability for the decayed
events, starting from un-decayed ones, is obtained from
\beq
 \label{eq:dec_prob}
 dP(\decPS | \undecPS) =\frac{1}{\mbox{BR}(t\to b \bar{\ell} \nu)
  \ \mbox{BR}(\tb \to \bar{b} \ell \bar{\nu})} \,
 \frac{\mathcal{M}_{\rm dec.}(\undecPS,\dectPS,\dectbarPS)}
 {\mathcal{M}_{\rm undec.}(\undecPS)} \, d\dectPS \, d\dectbarPS 
 \, ,
\eeq 
where $\decPS=\{\undecPS,\dectPS,\dectbarPS\}$ is the full space phase including the decay products,
while $\mathcal{M}$'s are the matrix elements squared for the given partonic process.  
In practice, however, to generate  $\decPS$ configurations according
to eq.~(\ref{eq:dec_prob}), one makes use of an upper bound $U_{\rm dec.}$ for the ratio
$\mathcal{M}_{\rm dec.} / \mathcal{M}_{\rm undec.} $ and then proceed as follows:
\begin{enumerate}[I.]
\item \label{item:veto} Generate a tentative decay kinematics $\decPS$, starting from $\undecPS$.
\item Extract a random value $r$   in the range $\lq 0,
U_{\rm dec.}  \rq$.
\item If $r< \mathcal{M}_{\rm dec.}(\decPS) / \mathcal{M}_{\rm
undec.}(\undecPS)$, then the decay kinematics is allowed and the event is accepted. Otherwise 
go back to step~\ref{item:veto}.
\end{enumerate}

An efficient functional form of the upper-bounding function may be readily reconstructed 
from the structure of top-quarks decay:
\begin{equation}
\label{eq:ub_dec}
U_{\rm dec.} (\decPS) = \mathcal{N} \
\frac{\mathcal{M}_{t\to b W}(M^2_t,M_{\bar{\ell} \nu}^2)}
{(M^2_t-m_t^2)^2 + m_t^2 \, \Gamma_t^2}
\ \frac{\mathcal{M}_{W\to \bar{\ell} \nu}(M_{\bar{\ell} \nu}^2)}
{(M_{\bar{\ell} \nu}^2-m_W^2)^2 + m_W^2 \, \Gamma_W^2}\, \times ( t \leftrightarrow \tb )\,,
\end{equation}
where $M^2_t$ and $M_{\bar{\ell} \nu}^2=(k_{\bar{\ell}}+k_{\nu})^2$ are
the top-quark and W boson virtualities and $\mathcal{M}_{t\to b W}$
and $\mathcal{M}_{W\to \bar{\ell} \nu}$ are the squared amplitudes for
the $1\to2 $ decay processes.  The choice of the normalization factor
$\mathcal{N}$ has been performed by sampling the allowed phase space,
by requiring that the inequality
\begin{equation}
\mathcal{M}^f_{\rm dec.}(\undecPS,\decPS) \le \mathcal{M}^f_{\rm
  undec.}(\undecPS)\ U_{\rm dec.}(\decPS)
\end{equation}
always holds, for any given specific subprocess flavour $f$.

In the rest of this section we present results obtained within
this approach.  We have set \beqn\nonumber &\Gamma_t=1.31~\rm GeV\,,
\quad m_W=80.398~\rm GeV\,, \quad \Gamma_W=2.141~\rm GeV\,,&\\ &
\aem=1/128.89\,, \quad \sin^2\theta_W=0.22265\,, & \eeqn and we have
considered a simplified form for the CKM matrix, with a mixing between the first two generations only: 
\beq V_{ud}=V_{cs}=0.975\,,\quad V_{us}=V_{cd}=0.222\,,\quad V_{tb}=1
\eeq and all the other entries set to zero. 
No extra acceptance cuts on the leptons are placed in the following plots.

\begin{figure}[tb]
\centering
\vspace*{10mm}
    {
    \includegraphics[width=\figwidth]{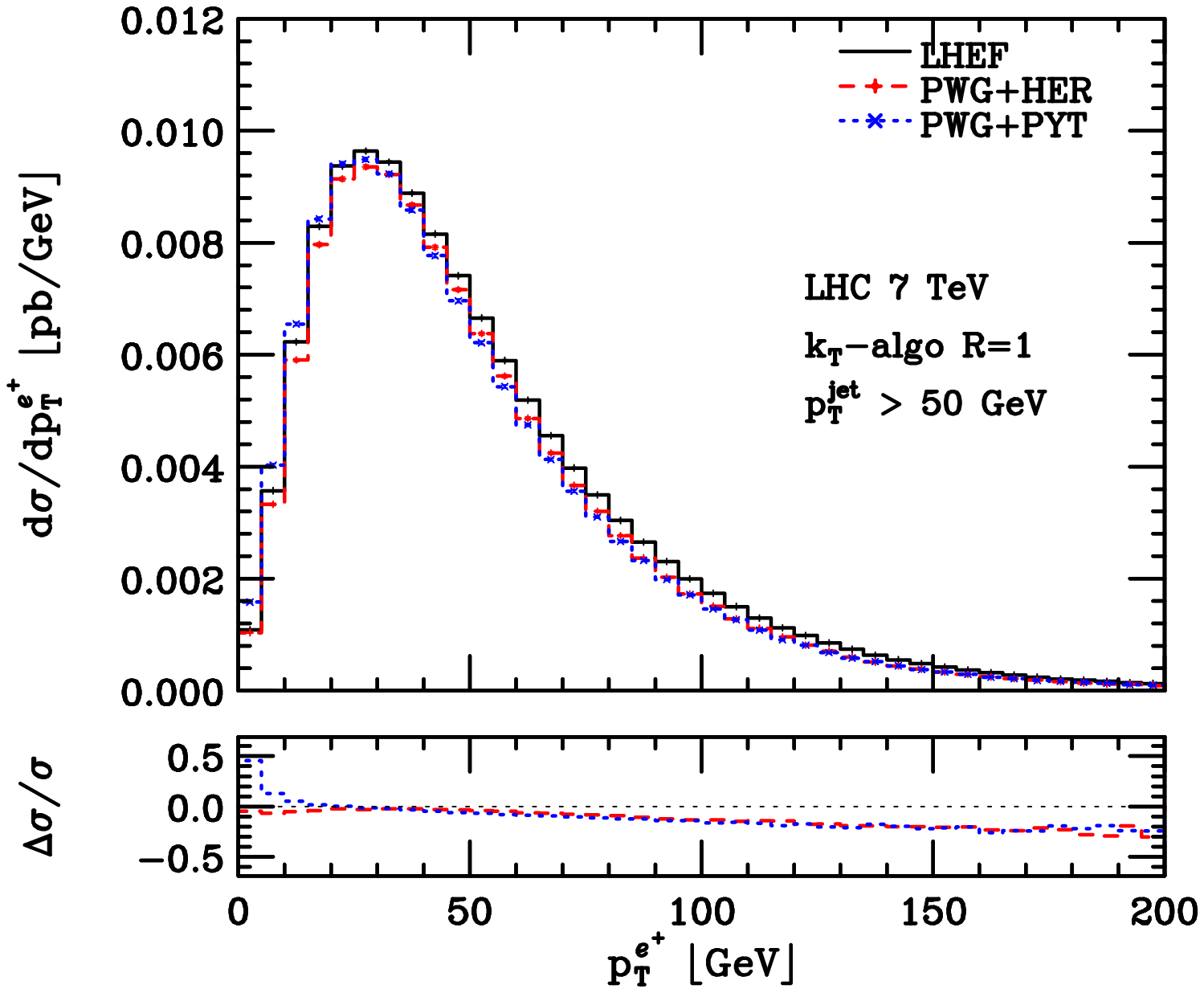}
    \hfill
    \includegraphics[width=\figwidth]{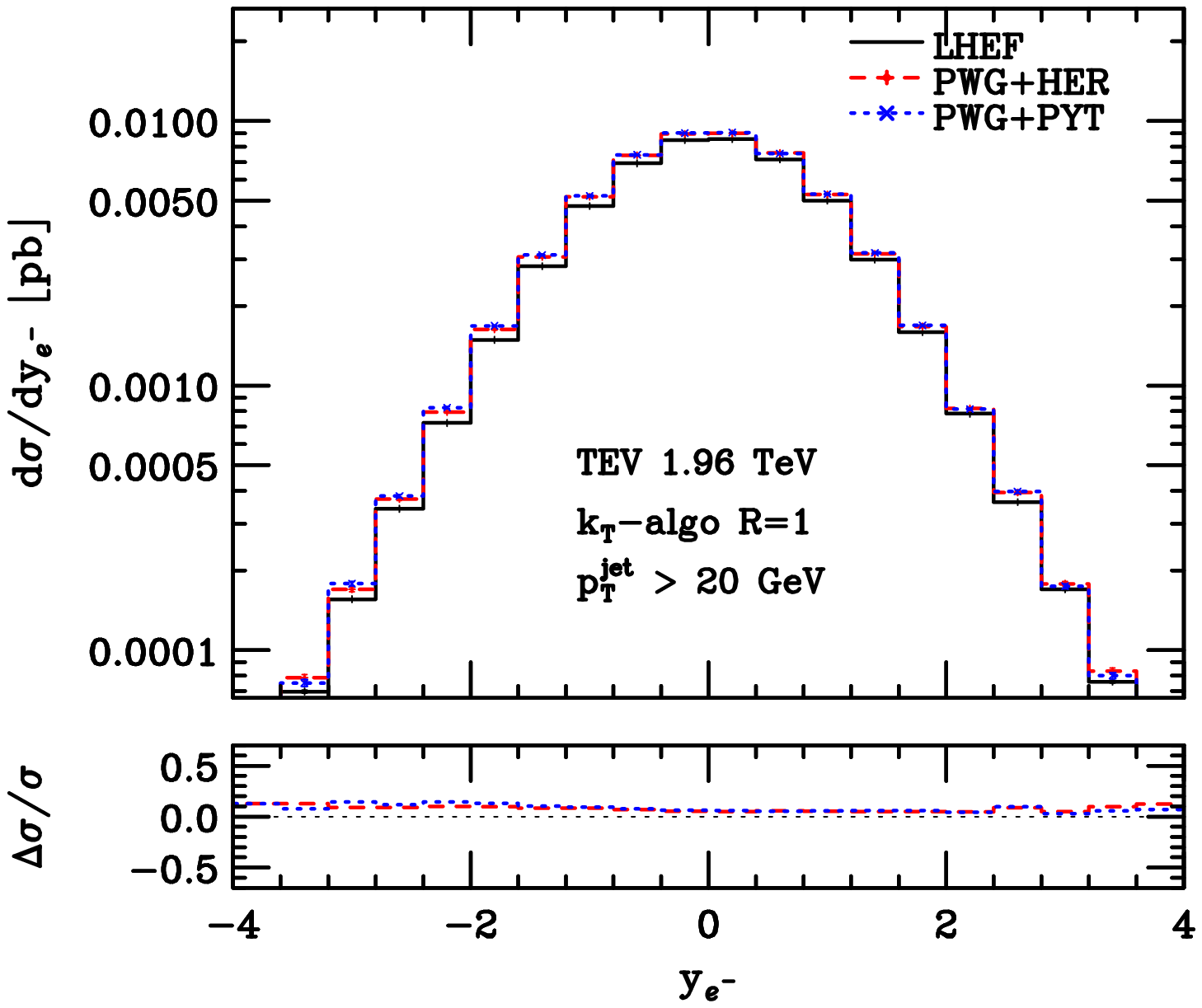}
    }
    \caption{ \small
      \label{fig:lep-TEV-LHC}
The differential cross sections as function of the positively charged lepton transverse momenta at the LHC ($\sqrt{s}=7$~TeV) and the the differential cross sections as function of the negatively charged lepton rapidity at the Tevatron ($\sqrt{s}=1.96$~TeV). }
\end{figure}

With these ingredients we are able to offer predictions for top-quarks
decay products, at the LHEF level and after shower and hadronization.
We compare them in Fig.~\ref{fig:lep-TEV-LHC} , where we plot the
positively charged lepton transverse momentum and the rapidity of the
negatively charged one, at the $7~\rm TeV$ LHC and $1.96~\rm TeV$
Tevatron colliders, respectively.  We note that in these distributions
 shower effects are rather small. Moreover, one does not expect to
see any dependence on the top-quark spin.

Spin correlation effects in $\ttb $ production are usually studied
by choosing $t$ and $\tb $ spin quantization axis and looking at
reference frames and angular distributions that are particularly
sensitive to these correlations\footnote{See {\it e.g.},~\cite{Uwer:2004vp}
for strategies to maximize the spin correlation of $\ttb$-pairs produced at the LHC.},
a procedure that can easily be performed in SMC simulations. 
However, due to the presence of unobserved neutrinos, this can become a difficult task 
in the data analysis for an experimental collaboration.  
An example of such observables is the double differential distribution, 
\beq
\label{eq:dcos1dcos2}
\frac{1}{\sigma} \frac{d^2\sigma}{d\cos \theta_1 d\cos \theta_2}\, ,
\eeq
 where  $\theta_1$ and $\theta_2$ are the angles 
between the directions of the flights of the leptons coming from the
decayed top-quark in the $t$ ($\tb $) rest frame and an arbitrary
direction that defines the quantization axis for the (anti-)top-quark spin.

The double differential distribution in eq.~(\ref{eq:dcos1dcos2}) can
be evaluated to good approximation to
\beq
\label{eq:dcos1dcos2approx}
 \frac{1}{4} \left( 1 - \kappa\, \cos \theta_1 \cos \theta_2\right)\,,
 \eeq which is only valid, though, for fully inclusive cross sections.
 As soon as cuts are imposed the form of the distribution will change
 and a further dependence on $\cos \theta_{1,2}$ could
   be introduced.\footnote{We note
   that electroweak effects and QCD absorptive parts lead to a tiny
   polarization of the top-quark and thus a deviation from the simple
   form in eq.~(\ref{eq:dcos1dcos2approx}) of the double differential
   distribution shown in eq.~(\ref{eq:dcos1dcos2}), even in the
   absence of cuts.}  The cuts may also affect the simple
 interpretation of $\kappa$, which in general depends also on the
 choice of the spin quantization axis. Nevertheless even in the
 presence of cuts the distribution is still a useful observable to
 study the impact of spin correlations. This quantity has been
 recently measured by D0~\cite{Abazov:2011qu} for the choice of the
 beam direction as quantization axis, finding $\kappa = 0.10 \pm
 0.45$. Due to the large uncertainties, this value is both compatible
 with no spin correlation and with the NLO predictions $\kappa =
 0.777$ of~\cite{Bernreuther:2001rq}.

\begin{figure}[tb]
\centering
\vspace*{10mm}
    {
      \includegraphics[width=\figwidth]{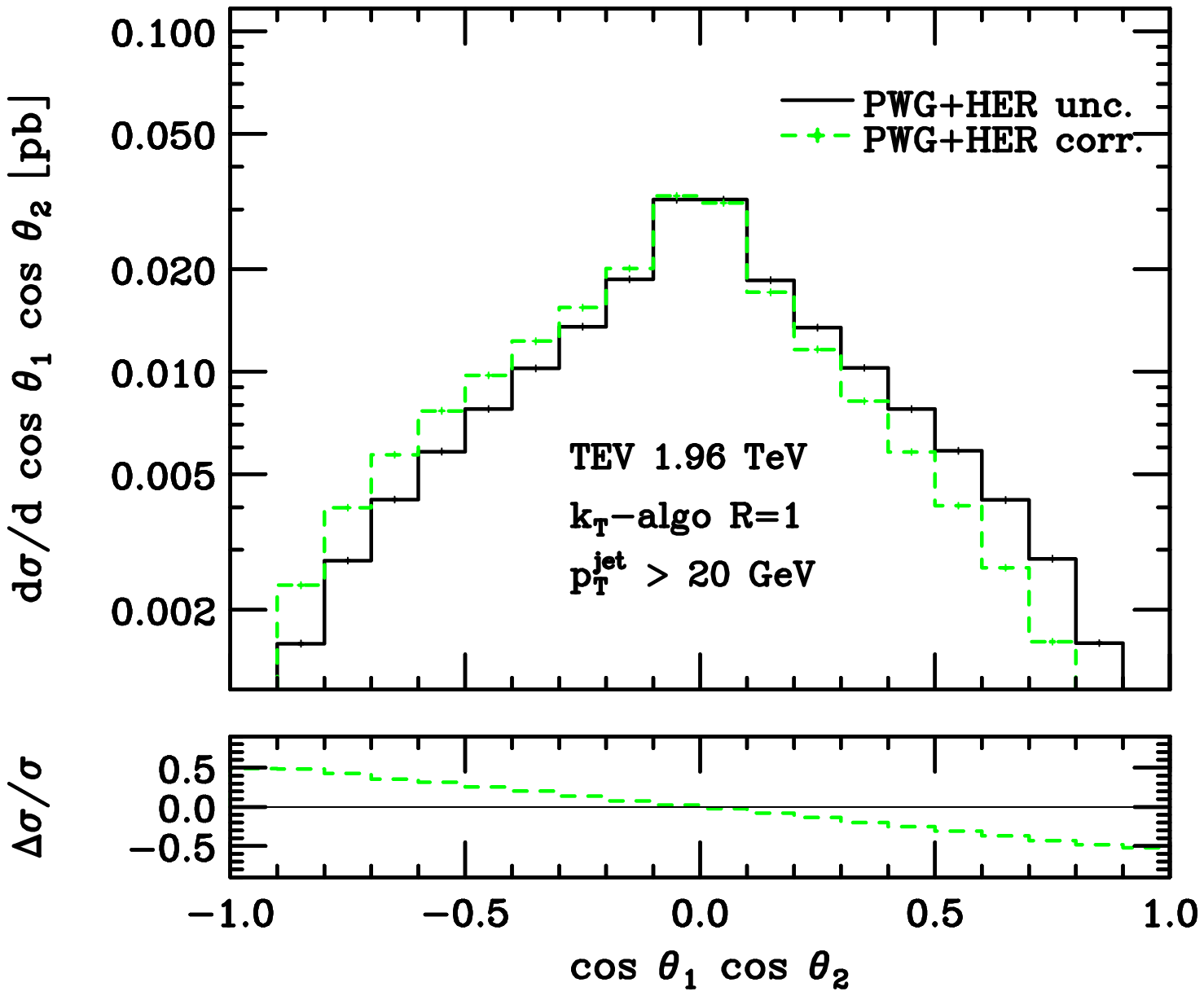}
    \hfil
    \includegraphics[width=\figwidth]{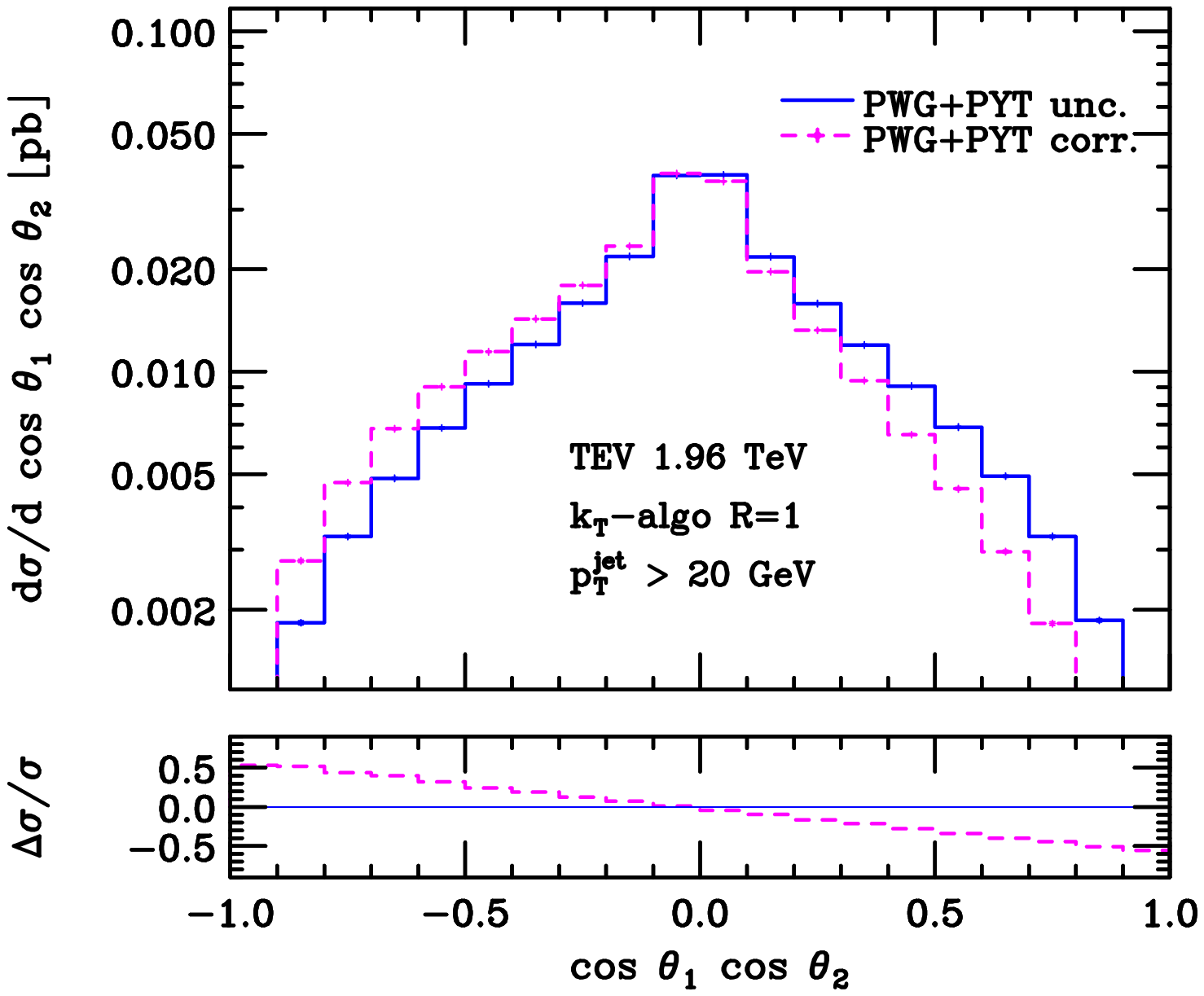}
    }
    \caption{ \small
      \label{fig:costh1costh2} 
     Effect of the inclusion of spin correlations  when interfacing to \HERWIG{} (\figleft{}) and \PYTHIA{} (\figright{}) programs. 
     Results for the Tevatron collider ($\sqrt{s}=1.96$~TeV).  } 
\end{figure}

\begin{figure}[tb]
\centering
\vspace*{10mm}
    {
    \includegraphics[width=\figwidth]{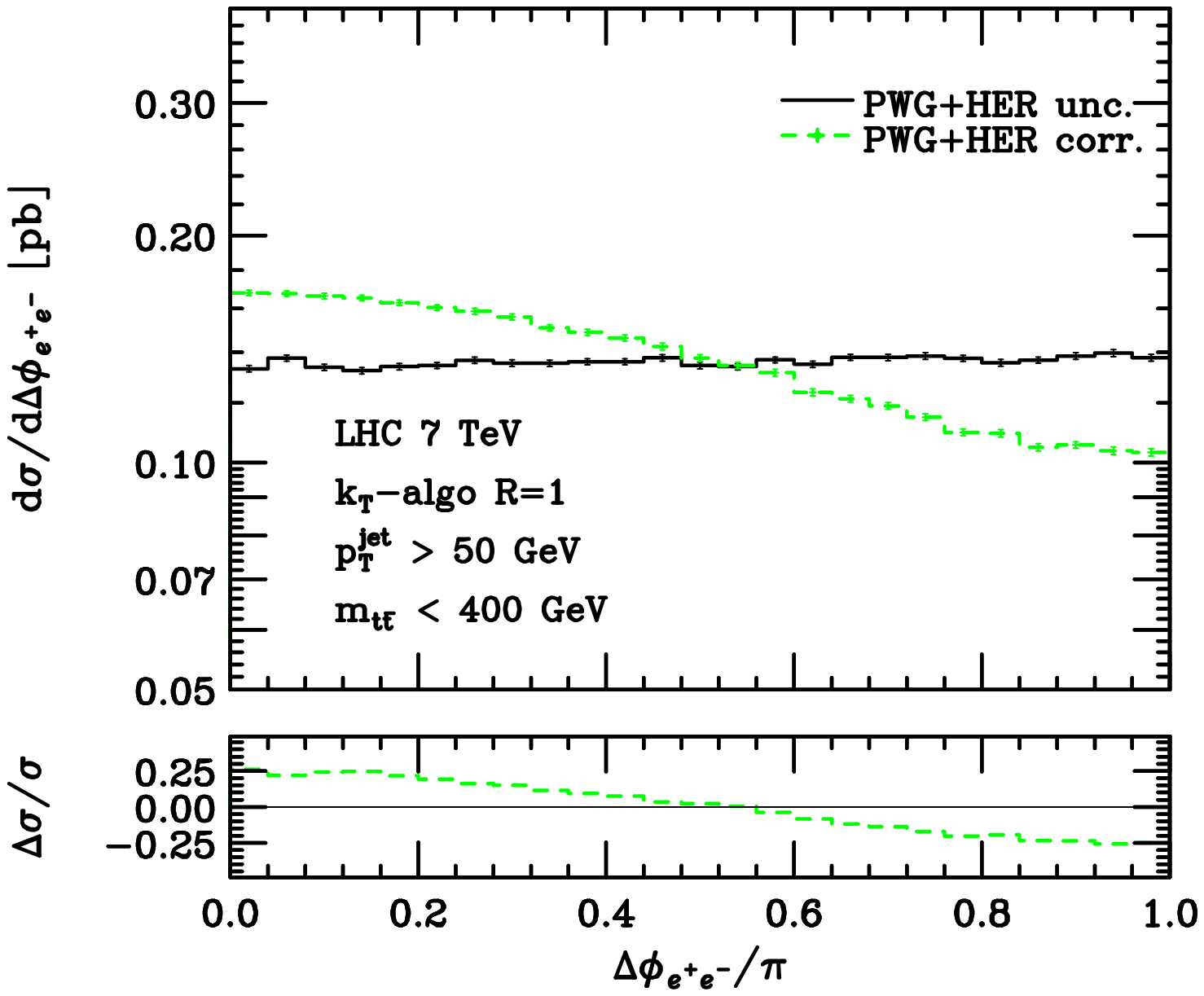}
    \hfil
    \includegraphics[width=\figwidth]{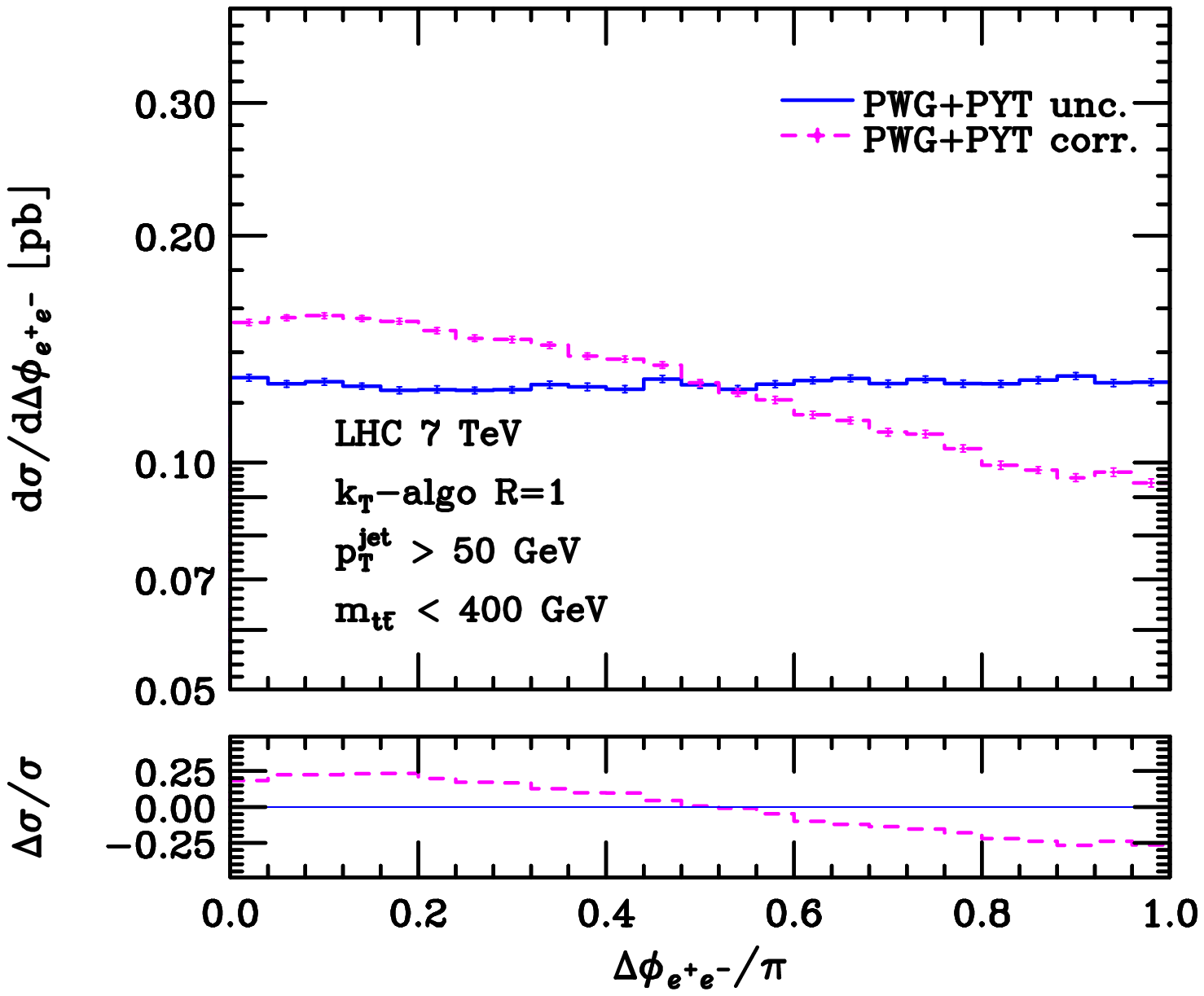}
    }
    \caption{ \small
      \label{fig:Dphimtt400} 
     Effect of the inclusion of spin correlations when interfacing to \HERWIG{} (\figleft{}) and \PYTHIA{} (\figright{}) programs. 
     Results for the LHC collider ($\sqrt{s}=7$~TeV).  } 
\end{figure}

\clearpage
 In
Fig.~\ref{fig:costh1costh2} we present our predictions for the
distribution in eq.~(\ref{eq:dcos1dcos2}) at the Tevatron using the
beam-axis to define the lepton angles, obtained by including the spin
correlation effects as explained above in our $\ttb j $ sample.  We
compare these results with those obtained by letting the SMC program
performing the uncorrelated top-quark decays.  The effect of the spin
correlations is clearly visible, leading to differences as large as
$50\%$ at extreme $\cos \theta_1 \cos \theta_2 $ values.

Alternatively, one can look at the distributions of the top-quark decay products in the laboratory frame.  In this case the problem is to devise clever enough cuts 
to single out spin-correlation effects which are usually hardly visible.\footnote{
A recent paper~\cite{Melnikov:2011ai} suggests the use of these distributions to construct a
likelihood function for the spin-correlation hypothesis, 
that can be then tested statistically for its validity. See also  \cite{Abazov:2011ka,Abazov:2011gi} for up-to-date results using this approach at Tevatron.}

There are, however, notable exceptions.  One such example, the
azimuthal distance between the leptons coming from the top- and the
anti-top-quark, has already been shown in~\cite{Mangano:2006rw} to
depend mildly on spin-correlations. Results for this observable has
also been presented in~\cite{ATLAS-CONF-2011-117}, for the ongoing LHC
$7$ TeV run.  Later on, it has been suggested that imposing the
$m_{\ttb }< 400~\rm{GeV} $ acceptance cut should enhance this
dependence in the di-lepton search channel at the
LHC~\cite{Mahlon:2010gw}.  However, due to the inherent ambiguities of
the unobserved neutrinos, the true $m_{\ttb }$ value cannot be
reconstructed on an event-by-event basis in a real experiment.  In our
case, we simply looked back in the shower history to reconstruct
unambiguously the true $m_{\ttb }$.  A more realistic study would use
the definition of $m_{\ttb }$ value averaged over all the possible
neutrino assignments proposed by the authors of~\cite{Mahlon:2010gw}.
However, as shown in~\cite{Mahlon:2010gw}, this only leads to minor
differences.

 In Fig.~\ref{fig:Dphimtt400} we show our results with
both the \HERWIG{} and \PYTHIA{} showers and the hadronization stage
simulations. In both panels, spin correlation effects are again
clearly visible as a change in the slope of the distribution.

As a final remark, we investigate the effects induced by top-quark
charge asymmetries on the di-lepton final state, by studying the
leptonic charge asymmetries.  This issue has also been addressed in a
recent paper~\cite{Bernreuther:2010ny}, for the $t \bar{t}$ inclusive
production, including next-to-leading order QCD and EW corrections.
In ref.~\cite{Melnikov:2010iu}, the authors provide predictions for
this quantity at NLO in the $\ttb j$ sample.  The definitions for the
asymmetries follow the ones in eq.~(\ref{eq:afbt}), but are expressed
in term of leptonic rapidities in the laboratory frame:
\beq
\label{eq:afbep}
A^{\ell^+}_{\rm FB} = \frac{1}{\sigma} \left( \quad {\displaystyle
  \int\limits_{\scriptscriptstyle y_{\ell^+}>0}} d\sigma -
{\displaystyle \int\limits_{\scriptscriptstyle y_{\ell^+}<0}} d\sigma
\ \right) \,,\qquad A^{\ell^+ \ell^-}_{\rm FB} = \frac{1}{\sigma}
\left( \quad {\displaystyle \int\limits_{\scriptscriptstyle \Delta
    y_{\ell^+ \ell^-}>0}} d\sigma - {\displaystyle
  \int\limits_{\scriptscriptstyle \Delta y_{\ell^+ \ell^-}<0}}
d\sigma\ \right)\,, \eeq with $\Delta y_{\ell^+ \ell^-}=
y_{\ell^+}-y_{\ell^-}$.  We present our results in
Tab.~\ref{tab:yasym3}, also reporting, for ease of comparisons, the
results including UE and MPI and the ones for the uncorrelated case,
which we dub PWG+HER' and PWG+PYT', where the (anti-)top-quark decay
is performed by the SMC neglecting spin-correlation effects .
Comparing the results including the spin correlation to the results
where the top-quark decay is handled by the shower program --- spin
correlations are thus not taken into account in the latter
predictions--- we observe only a small difference between the two
approaches.  On the other hand a significant change can be observed
when comparing LHEF results with the results including shower effects.
Our conclusion is that most of the effect present at parton level is
washed out by the parton shower. Indeed, investigating the behaviour
of the leptonic forward-backward charge asymmetries in presence of a
cut on the transverse momentum of the intermediate $\ttb$ pair, one
can trace back the large differences between results at partonic and
hadronic level to the region of low $\pt^{\ttb}$'s once again, as it
was the case for the top-quark forward-backward charge asymmetry shown
in Tab.~\ref{tab:yasym}.

\section{Conclusions}
\label{sec:conclusions}

We have studied top-quark hadroproduction in association with a jet
merged with parton showers to NLO in QCD.  For that purpose, we have
presented an implementation of the process $\ttb +$jet in the
framework of \POWHEGBOX{}. A careful validation has been performed by
comparing with known results.  For Tevatron and LHC at
$\sqrt{s}=7~$~TeV we have investigated a large number of
distributions, some of them with particular sensitivity to parton
shower effects, such as the differential cross sections as function of
the transverse momentum of the $\ttb$-pair and the hardest jet.  We
have presented a detailed study of the impact of the parton-shower on
the top-quark charge asymmetry. We find that the inclusion of the
shower changes significantly the NLO predictions. A detailed analysis
allowed us to trace back the origin of these corrections to the low
$p_T^{\ttb}$ region. We stress that the results presented here
represent the first NLO evaluation of this distribution. Excluding the
region $p_T^{\ttb}<10$~GeV we find that the parton shower leads only to a
marginal change of the charge asymmetry binned in $p_T^{\ttb}$.  A
detailed comparison with future measurements may help in clarifying
the discrepancy observed in the inclusive forward-backward charge
asymmetry.  However, we stress that in this work, we always refer to
the asymmetries defined for the $\ttb j$ sub-sample of the inclusive
$\ttb$ production.

We have also taken into account top-quark decays at LO in QCD, so that
spin correlations, {\it e.g.}, for the leptons from the decaying
$\ttb$-pair can be studied. Similar to the case of inclusive top-quark
pair production the effects of spin correlations are clearly visible
in double differential distributions of the top-quark (and anti-quark)
decay products. Since $\ttb +$jet  events represent an
important fraction of all $\ttb$ events, the measurement of this
effect should be feasible at Tevatron and LHC.

The present work offers several options for extension.  For instance,
we have restricted ourselves in this article entirely to the study of
top-quark hadroproduction, although the available QCD corrections at
NLO~\cite{Dittmaier:2007wz,Dittmaier:2008uj} are equally applicable to
the process $b{\bar b}+$jet with massive bottom-quarks.  Therefore,
the investigation of $b{\bar b}+$jet in the framework of \POWHEGBOX{}
is feasible, even if further refinements in the code will be
required to perform this task.

Most important, however, will be an extension of the phenomenological
analysis presented in this paper to meet the real conditions of the
experimental environment, {\it e.g.}, studying more carefully tuning,
UE and MPI effects of the SMC and, potentially, for other acceptance
cuts.  Also, results with different values of the top-quark mass,
strong coupling constant $\as(M_Z)$, $\mur$, $\muf$ scales and with
modern sets of PDFs will be needed.

Another natural development is the application of the merging
procedure recently proposed in~\cite{Alioli:2011nr} to the \MENLOPS{}
improved $\ttb $ and $\ttb j$ samples, in order to improve the
description of both fully inclusive $\ttb $ and inclusive $\ttb +$jet
productions.  However, this will be the subject of a separate
publication.

In general, we believe that a more thoroughly phenomenological study 
would be better performed in the framework of an experimental
collaboration. Therefore, in order to facilitate all these future
studies, the code of the \POWHEGBOX{} implementation for the process
$\ttb +$jet will be made publicly available on the web-page {\tt
  http://powhegbox.mib.infn.it}.

\subsection*{Acknowledgments}
We are thankful to M.~Garzelli, E.~Re and Z.~Trocsanyi for useful discussions. 
This work has been supported by Helmholtz Gemeinschaft under contract VH-HA-101
({\it Alliance Physics at the Tera\-scale}), by Deutsche
Forschungsgemeinschaft in Sonderforschungsbereich/Transregio~9 and by
the European Commission through contract PITN-GA-2010-264564 ({\it
  LHCPhenoNet}).

\appendix
\section{Folding}
\label{app:folding} In this appendix we briefly review the folding procedure, its
implementation in the \POWHEGBOX{} and the application to the present calculation.
For a more detailed explanation of the procedure itself we refer to~\cite{Alioli:2010qp}.
\begin{table}[!hbt]
\begin{center}

\begin{tabular}{ccccccc}
\hline
\hline
$\xi$-$y$-$\phi$ folding $\Longrightarrow$&  1-1-1 &  5-1-1 & 1-5-1 & 1-1-5 & 5-5-5 &  5-5-10 \\
\hline
\hline
TeV  1.96 TeV, $\pt^{\rm gen}=2$~GeV  & 0.20 & 0.17 & 0.14 & 0.16 &  0.07 & 0.06 \\
LHC 7 TeV,  $\pt^{\rm gen}=5$~GeV& 0.22 & 0.19 & 0.16 & 0.19 & 0.12 &  0.11\\
LHC 14 TeV,  $\pt^{\rm gen}=5$~GeV & 0.28 & 0.23 & 0.26 & 0.19 & 0.17 &  0.16\\
\hline
\hline

\end{tabular}
\end{center}
\caption{
\label{tab:folding}
The fraction negative weighted events generated using different folding factors. 
}
\end{table}

The \POWHEG{} method allows for the generation of positive weighted
events only.  This is a direct consequence of the perturbative nature
of the exclusive differential cross section integrated over the Born
variables $\bar{B}$ (see {\it e.g.}, eq.~4.2 in Sec.~4
of~\cite{Alioli:2010xd} and comments thereafter).  In our
implementations this integration is performed in the Monte Carlo
approach: we generate fully exclusive configurations which depend also
on the three variables associated to the extra radiation ($\xi,y,\phi$
in the FKS approach~\cite{Frixione:1995ms}) and then we sum over them
discarding the extra dependency.  However, the weight of this fully
exclusive configurations, which is dubbed $\tilde{B}$, is not
guaranteed to be always positive, in its enlarged phase space.
The positivity of $\tilde{B}$ may always be recovered by
keeping the Born variables fixed and redefining $\tilde{B}$ as the average
over more phase space points, which differs only in the radiative
variables. This clearly corresponds again to integrating over the
radiative variables. This is the essence of the {\it folding}
procedure: the number of phase space points considered in each
evaluation of the $\tilde{B}$, for each radiative variable, is called
{\it folding factor}.

An immediate drawback is that the folding procedure results in a
increase of the program run time which is proportional to the product
of the folding factors for the three radiation  variables. In
particular cases, it turns out to be more practical including negative
weighted events in the final sample.  However, these performance costs
in the generation stage may be well balanced if full detector
simulations are included or if it is required to have positive weights
only. For this reasons, we report in Tab.~\ref{tab:folding} the
observed percent occurrence of negative weighted events in case of
different folding levels and different colliders.

{\small
\bibliographystyle{JHEP}
\bibliography{paper}
}

\end{document}